\documentclass[lettersize,journal,onecolumn]{IEEEtran}
\usepackage[T1]{fontenc}
\usepackage{amsmath,amssymb,amsfonts}
\usepackage{algorithmic,algorithm}
\usepackage{array,tabularx,booktabs}
\usepackage{graphicx}
\usepackage{textcomp,stfloats,url,cite}
\usepackage[hidelinks,hypertexnames=false]{hyperref}
\newcolumntype{Y}{>{\raggedright\arraybackslash}X}

\newsavebox{\equationbox}
\newcommand{\fitmath}[1]{%
  \sbox{\equationbox}{$\displaystyle #1$}%
  \ifdim\wd\equationbox>.88\columnwidth
    \resizebox{.88\columnwidth}{!}{\usebox{\equationbox}}%
  \else\usebox{\equationbox}\fi}
\begin{document}
\title{Residual-Closed-Loop SAGE--Brent Estimation for Unknown-Order Wideband Underwater Acoustic Channels}

\author{Wenxu~Bi, Peng~Chen, and Wei~Wang\thanks{All authors are with the School of Information Engineering, Chang'an University, Xi'an 710064, China. Corresponding author: Peng Chen. E-mail: Wenxu~Bi (2026024009@chd.edu.cn), Peng~Chen (pchen@chd.edu.cn), Wei~Wang (wei.wang@chd.edu.cn).}}
\maketitle

\begin{abstract}

Grid mismatch in wideband channel estimation leaves structured cancellation residuals after strong-path subtraction. These residuals generate spurious correlation peaks or mask weak arrivals, turning continuous parameter errors into path-count errors. We propose RC-SAGE--Brent, a residual-closed-loop estimator that jointly refines continuous delay--scale parameters and corrects the unknown-order path set. Within each feedback cycle, ambiguity-constrained alternating Brent searches suppress strong-path cancellation residuals. Candidate validation, resolution-constrained pruning, and full-domain residual redetection then revise the active path set. For linear frequency modulation (LFM) probes, an observable signal-to-residual ratio adaptively scales the model-order penalty to mitigate overestimation caused by delay--scale coupling. Theoretical analysis links off-grid parameter errors to local cancellation-residual curvature and weak-path detectability. Factorial ablations, paired comparisons against standard sparse and off-grid estimators, and measured channel replay demonstrate improved path-set recovery and waveform reconstruction while delineating the operational boundaries of the method.

\end{abstract}

\begin{IEEEkeywords}

Underwater acoustic channel estimation, residual feedback loop, continuous SAGE, Brent search, off-grid estimation, unknown model order.

\end{IEEEkeywords}

\section{Introduction}\label{sec:1}

Underwater acoustic channels exhibit low sound speeds, extended delay spreads, and rapid time variation. A typical received signal comprises a few dominant propagation paths superimposed on weaker scattered arrivals. Under wideband conditions, time scaling characterizes Doppler effects more accurately than narrowband frequency shifts. Both path delays and Doppler scales take continuous physical values. Researchers have applied sparse channel estimation to multicarrier underwater communications and time-varying channel tracking \cite{Berger2010SparseUWA,LiPreisig2007RapidUWA,Jiang2011DelayDoppler,vanWalree2013UWB,Stojanovic2008OFDM}. However, restricting continuous parameters to a discrete search grid inevitably introduces mismatch between candidate dictionary waveforms and true path arrivals.

Grid mismatch degrades performance beyond the familiar quantization error floor at high signal-to-noise ratios (SNRs). When a strong path falls between grid points, a discrete dictionary cannot synthesize its exact waveform. Even when the complex gain is reestimated optimally, subtracting the mismatched reconstruction leaves structured energy. We define this unmodeled component as the cancellation residual of the path. This cancellation residual corrupts the observation residual alongside unestimated paths and noise. Because its structure follows the ambiguity function of the probe near the true parameters, a strong-path cancellation residual can create spurious correlation peaks or mask genuine weak paths in near--far environments. Consequently, conventional one-shot order estimation frequently fails. Conversely, when the estimator reconstructs and cancels the strong path with continuous-parameter accuracy, rescanning the observation residual reveals weak paths that were previously obscured.

Existing algorithms address continuous-parameter and sparse estimation from different perspectives. Space-alternating generalized expectation-maximization (SAGE) decomposes joint multisource optimization into sequential single-path updates \cite{FesslerHero1994SAGE}. RELAX and WRELAX improve parameter accuracy by cyclically reestimating individual components \cite{LiStoica1996RELAX,LiWu1998WRELAX}. Orthogonal matching pursuit (OMP) provides low-complexity sparse recovery by greedily selecting the strongest correlation peak \cite{TroppGilbert2007OMP}, whereas Newtonized OMP (NOMP) extends frequency estimation to the continuum through Newton refinements after coarse detection \cite{Mamandipoor2016NOMP}. In Bayesian formulations, sparse Bayesian learning (SBL) and off-grid SBL employ hierarchical priors and grid perturbations to capture parameter uncertainty \cite{Tipping2001SBL,YangXieZhang2013OGSBL}. Yang and Choo recently formulated matched-filtered time-domain observations as a multiple-measurement-vector off-grid SBL model with first-order delay perturbations, demonstrating arrival resolution on SAVEX15 shallow-water data \cite{YangChoo2025TimeDomainOGSBL}. However, their model targets repeated observations across time and does not estimate wideband Doppler scale. In underwater communications, deep learning has been explored for channels with uncertain sparsity \cite{FengEtAl2024UncertainSparsity}, and off-grid OMP with maximum-likelihood refinement has been developed for underwater orthogonal time frequency space (OTFS) modulation \cite{YangEtAl2026UWAOTFS}. For detection thresholding, constant false-alarm rate (CFAR) rules scale thresholds dynamically with the estimated residual noise floor \cite{Rohling1983CFAR}.

Recently, Pinto and Juntti proposed robust OFDM-SAGE for parametric orthogonal frequency-division multiplexing (OFDM) channels \cite{PintoJuntti2026RobustOFDMSAGE}. Their algorithm iteratively updates multidimensional harmonic parameters, initializes new paths, and selects model order. This framework demonstrates the benefit of combining continuous local optimization with path discovery. Wideband underwater channels, however, introduce severe probe-dependent delay--scale coupling, high ambiguity sidelobes, and extreme near--far dynamics. Standard sequential estimators use the residual only in a forward pass to select the next candidate or terminate the search. If an early iteration admits a spurious peak or misses a weak arrival, that error permanently biases subsequent residuals. Local parameter refinement alone cannot remove false paths or recover missed ones. Therefore, an effective estimator must integrate continuous parameter refinement, cancellation-residual suppression, and path-set revision into a bidirectional feedback loop.

We propose residual-closed-loop SAGE--Brent (RC-SAGE--Brent) to close this loop. The estimator treats the observation residual as a dynamic feedback state and the active path set as the model state under revision. Continuous alternating SAGE--Brent searches refine delays and Doppler scales to suppress strong-path cancellation residuals. Candidate validation rejects noise peaks, resolution-constrained pruning removes duplicate components, and conditional least-squares refitting updates path gains. Rescanning the updated residual across the full parameter domain then recovers weak paths that were previously masked. Every structural change to the path set triggers a new cycle of continuous refinement and cancellation. For linear frequency modulation (LFM) probes, we also introduce a residual-adaptive complexity penalty that suppresses spurious detections along delay--scale ambiguity ridges.

First, we establish a residual feedback mechanism for unknown-order path correction. The algorithm recomputes the observation residual at each outer cycle and integrates candidate validation, resolution pruning, and full-domain redetection to purge spurious components and recover masked weak paths. Second, we design an ambiguity-constrained SAGE--Brent refinement strategy. Alternating one-dimensional Brent searches replace costly two-dimensional fine-grid evaluations, while ambiguity main-lobe widths and adjacent-path midpoints bound local search intervals to prevent path collisions. For LFM waveforms, an observable signal-to-residual ratio adaptively scales the model-order penalty without extra matrix inversions. Third, we validate the framework through factorial ablations, paired Monte Carlo benchmarks against OMP, WRELAX, SBL, and two-dimensional NOMP, and measured channel replay, while explicitly documenting computational costs and operational limits.

\section{Signal Model and Problem Definition}\label{sec:2}

\subsection{Continuous Wideband Multipath Model}\label{sec:2.1}

Let \(s(t)\) denote the transmitted complex analytic bandpass signal, \(T\) denote the observation duration, and \(T_s=1/f_s\) denote the sampling interval. Our synthetic experiments directly generate and estimate time-scaled waveforms in the sampled analytic bandpass domain, typically within a 6--10 kHz probe band. Within a locally stationary observation window, we model the continuous wideband multipath received signal as

\begin{equation}\tag{1}\label{eq:1}
\fitmath{\begin{gathered}
y(t)=\sum_{\ell=1}^{L}\alpha_\ell
s\!\left((1+\epsilon_\ell)(t-\tau_\ell)\right)+w(t),
\\
0\le t<T ,
\end{gathered}}
\end{equation}

where \(L\) is the path count; \(\alpha_\ell\in\mathbb C\), \(\tau_\ell\), and \(\epsilon_\ell\) are the complex gain, delay, and wideband Doppler scale of path \(\ell\), respectively; and \(w(t)\) is additive noise. We define the time-scaling factor as \(1+\epsilon_\ell\), where a positive scale \(\epsilon_\ell\) corresponds to temporal compression and an upward frequency shift. This definition remains independent of any radial-velocity sign convention. Let

\begin{equation}\tag{2}\label{eq:2}
\fitmath{\begin{gathered}
\mathbf a(\tau,\epsilon)
=\left[s\!\left((1+\epsilon)(nT_s-\tau)\right)\right]_{n=0}^{N-1},\\
\boldsymbol\theta=(\tau,\epsilon)^{\mathsf T},
\end{gathered}}
\end{equation}

denote the sampled single-path response vector and parameter pair. Collecting all \(N\) samples yields the discrete observation model

\begin{equation}\tag{3}\label{eq:3}
\fitmath{\mathbf y=\sum_{\ell=1}^{L}\alpha_\ell
\mathbf a(\boldsymbol\theta_\ell)+\mathbf w
=\mathbf A(\boldsymbol\Theta)\boldsymbol\alpha+\mathbf w,}
\end{equation}

with circularly symmetric complex Gaussian noise \(\mathbf w\sim\mathcal{CN}(\mathbf0,\sigma^2\mathbf I)\) in synthetic trials. We define the experimental SNR as the ratio of average noiseless response energy to noise variance,

\begin{equation}\tag{4}\label{eq:4}
\fitmath{\begin{gathered}
\mathrm{SNR}=10\log_{10}
\frac{\left\|\mathbf x\right\|_2^2}{N\sigma^2},
\\
\mathbf x=\mathbf A(\boldsymbol\Theta)\boldsymbol\alpha .
\end{gathered}}
\end{equation}

We next define the normalized wideband ambiguity correlation between two parameter pairs as

\begin{equation}\tag{5}\label{eq:5}
\fitmath{\rho(\boldsymbol\theta,\boldsymbol\theta')=
\frac{\mathbf a(\boldsymbol\theta)^{\mathsf H}
\mathbf a(\boldsymbol\theta')}
{\|\mathbf a(\boldsymbol\theta)\|_2
 \|\mathbf a(\boldsymbol\theta')\|_2}.}
\end{equation}

We use~\eqref{eq:5} to analyze cancellation residuals and local objective curvature. To calibrate the algorithmic main-lobe widths from the actual probe waveform, we set \(\mathbf a_0=\mathbf a(0,0)\) and evaluate the peak-normalized ambiguity slices

\begin{equation}\tag{5a}\label{eq:5a}
\fitmath{\begin{gathered}
C_\tau(\Delta\tau)=
\frac{|\mathbf a_0^{\mathsf H}\mathbf a(\Delta\tau,0)|^2}
{\|\mathbf a_0\|_2^4},\\
C_\epsilon(\Delta\epsilon)=
\frac{|\mathbf a_0^{\mathsf H}\mathbf a(0,\Delta\epsilon)|^2}
{\|\mathbf a_0\|_2^4}.
\end{gathered}}
\end{equation}

The calibration window and zero padding match those used to compute the waveform ambiguity function. We locate the first positive offset where each slice drops to \(10^{-3/10}\) via linear interpolation and double that offset to obtain the \ensuremath{-}3 dB full main-lobe widths \(W_\tau\) and \(W_\epsilon\). Whereas~\eqref{eq:5} normalizes each response vector to unit norm,~\eqref{eq:5a} normalizes solely by the zero-offset peak energy \(\|\mathbf a_0\|_2^4\). We use \(W_\tau\) and \(W_\epsilon\) to normalize parameter matching distances and bound local search intervals. Finally, we define the two-path near--far ratio (NFR) as

\begin{equation}\tag{6}\label{eq:6}
\fitmath{\Gamma_{\mathrm{NF}}=20\log_{10}
\frac{|\alpha_{\mathrm{s}}|}{|\alpha_{\mathrm{w}}|},}
\end{equation}

where the subscripts \(\mathrm{s}\) and \(\mathrm{w}\) denote the strong and weak paths, respectively.

\begin{figure*}[!t]
\centering
\includegraphics[width=\textwidth,height=.72\textheight,keepaspectratio]{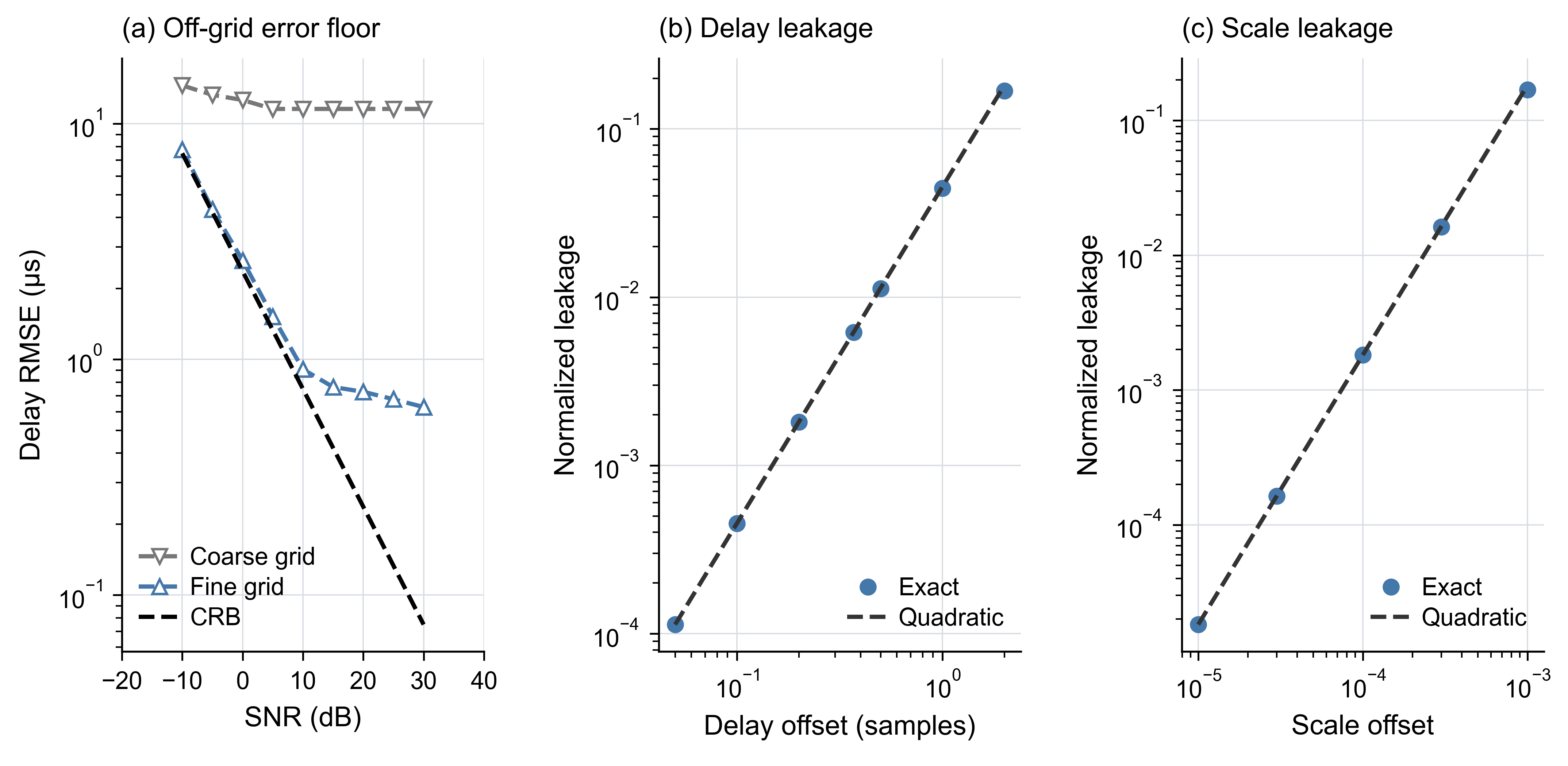}
\caption{Grid parameter mismatch and normalized cancellation-residual energy: (a) fixed-grid delay error floors; (b) effects of delay error; (c) effects of Doppler-scale error.}\label{fig:1}

\end{figure*}

Fig.~\ref{fig:1}(a) compares two fixed search grids against the Cram\'er--Rao bound (CRB) to show how quantization error floors emerge at high SNR. Figs.~\ref{fig:1}(b) and (c) sweep the delay and Doppler-scale offsets, comparing the exact normalized cancellation-residual energy against its local quadratic approximation. For small parameter offsets, the quadratic approximation closely tracks the exact residual energy. This agreement confirms that local ambiguity curvature governs the sensitivity of the cancellation residual to small off-grid errors. Larger offsets depart from the quadratic curve and require the exact correlation expression in~\eqref{eq:5}. Together, these curves show why continuous parameter refinement is necessary to suppress strong-path cancellation residuals. We quantify the resulting near--far detection gains in Fig.~\ref{fig:7} and Section~\ref{sec:7}.

\subsection{Estimation Tasks and Evaluation Metrics}\label{sec:2.2}

We evaluate the estimators under both known-order and unknown-order settings. In known-order estimation, the algorithm receives the true path count \(L\) and estimates the continuous parameter set \(\{\widehat\alpha_\ell,\widehat\tau_\ell,\widehat\epsilon_\ell\}_{\ell=1}^{L}\). This configuration isolates the benefits of continuous refinement under near--far interference and multipath coupling. In unknown-order estimation, the algorithm receives only an upper bound \(L_{\max}\) and jointly estimates the active path count \(\widehat L\le L_{\max}\) and the corresponding path parameters. An effective unknown-order estimator must simultaneously suppress noise false alarms, prune duplicate detections, and recover weak paths.

When two physical paths lie closer than the waveform resolution limit, estimating their individual complex gains becomes ill-conditioned and highly sensitive to noise. We therefore define resolvable effective paths according to the waveform resolution and evaluate path-detection metrics against these effective components.

Let \(\mathbf x\) and \(\widehat{\mathbf x}\) denote the true noiseless channel response and its reconstruction, respectively. We measure time-domain waveform reconstruction error via the normalized mean-squared error (NMSE),

\begin{equation}\tag{7}\label{eq:7}
\fitmath{\mathrm{NMSE}_{x}=10\log_{10}
\frac{\|\mathbf x-\widehat{\mathbf x}\|_2^2}
{\|\mathbf x\|_2^2}.}
\end{equation}

In measured channel replay experiments, we also compute the frequency-response NMSE across the active OFDM subcarrier set \(\mathcal K\),

\begin{equation}\tag{8}\label{eq:8}
\fitmath{\mathrm{NMSE}_{H}=10\log_{10}
\frac{\sum_{k\in\mathcal K}|H_k-\widehat H_k|^2}
{\sum_{k\in\mathcal K}|H_k|^2}.}
\end{equation}

When fitting measured impulse responses from the UWA Channel Library, we quantify the normalized noisy fitting residual as

\begin{equation}\tag{8a}\label{eq:8a}
\fitmath{E_{\mathrm{res}}=10\log_{10}
\frac{\|\mathbf y-\widehat{\mathbf x}\|_2^2}
{\|\mathbf y\|_2^2}.}
\end{equation}

Whereas~\eqref{eq:7} and~\eqref{eq:8} compare reconstructions against noiseless truth in simulations,~\eqref{eq:8a} evaluates the residual against the noisy observation \(\mathbf y\), reflecting both finite-path modeling error and background measurement noise.

To match estimated paths to true paths in two dimensions, we compute the normalized parameter distance

\begin{equation}\tag{9}\label{eq:9}
\fitmath{d_{ij}=\sqrt{
\left(\frac{\widehat\tau_j-\tau_i}{W_\tau}\right)^2+
\left(\frac{\widehat\epsilon_j-\epsilon_i}{W_\epsilon}\right)^2}.}
\end{equation}

An estimated path \(j\) matches a true path \(i\) if and only if \(|\widehat\tau_j-\tau_i|<0.5W_\tau\) and \(|\widehat\epsilon_j-\epsilon_i|<0.5W_\epsilon\). We omit the Doppler-scale term in one-dimensional delay-only evaluations. Let \(N_{\mathrm{TP}}\), \(N_{\mathrm{FP}}\), and \(N_{\mathrm{FN}}\) denote the numbers of true-positive, false-positive, and false-negative paths in a scenario, respectively. We compute precision \(P\), recall \(R\), and \(F_1\) score as

\begin{equation}\tag{10}\label{eq:10}
\fitmath{\begin{gathered}
P=\frac{N_{\mathrm{TP}}}{N_{\mathrm{TP}}+N_{\mathrm{FP}}},\\
R=\frac{N_{\mathrm{TP}}}{N_{\mathrm{TP}}+N_{\mathrm{FN}}},\\
F_1=\frac{2PR}{P+R}.
\end{gathered}}
\end{equation}

For empty or zero-detection boundary cases, we set \(P=R=F_1=1\) when \(L=0\) and \(\widehat L=0\); \(R=1\) and \(P=F_1=0\) when \(L=0\) and \(\widehat L>0\); and \(P=R=F_1=0\) when \(L>0\) and \(\widehat L=0\). Whenever \(P+R=0\) in nonempty trials, we set \(F_1=0\). Because \(\|\mathbf x\|_2=0\) in noise-only scenarios (\(L=0\)), response NMSE is undefined there; accordingly, we exclude noise-only scenarios from NMSE averages. In those noise-only runs, the nominal SNR simply sets the noise variance relative to a unit-gain reference path at \(\tau=5\ \mathrm{ms}\) and \(\epsilon=0\). We average \(P\), \(R\), and \(F_1\) across scenarios rather than pooling individual paths globally. Thus, overall \(F_1\) averages across all 1080 scenarios (including the 240 noise-only runs), whereas response NMSE averages across the 840 nonempty scenarios. Similarly, overall order accuracy is weighted by scenario count across the test suite.

In near--far diagnostic trials, we assess both parameter accuracy and amplitude fidelity for the weak arrival. For each designated weak path, we identify the closest parameter-matched estimate \(j_*\) within the resolution gate and set the position indicator \(D_{\mathrm{pos}}=1\). In Fig.~\ref{fig:7} and Fig.~\ref{fig:10}(a), we further require amplitude consistency within \(\pm6\ \mathrm{dB}\) of the true weak-path gain,

\begin{equation}\tag{10a}\label{eq:10a}
\fitmath{\begin{gathered}
D_{\mathrm{amp}}=
\mathbf1\!\left\{D_{\mathrm{pos}}=1\ \text{and}\quad
\left|20\log_{10}
\frac{|\widehat\alpha_{j_*}|}{|\alpha_{\mathrm w}|}\right|
\le6\ \mathrm{dB}\right\},\\
P_{\mathrm{amp}}=\frac1n\sum_{i=1}^n D_{\mathrm{amp},i}.
\end{gathered}}
\end{equation}

If no estimated path satisfies the parameter gate, we set \(D_{\mathrm{amp}}=0\). We use \(D_{\mathrm{amp}}\) strictly as an offline evaluation metric; the estimator itself never accesses true path amplitudes during candidate validation. In control trials without a weak path, we record any parameter-matched detection inside the designated gate as a false alarm.

Finally, we measure order estimation error through the absolute error \(|\widehat L-L|\) and its scenario average, the order mean absolute error (MAE). In unknown-order benchmarks, we flag a severe estimation failure whenever a trial misses at least half of the true paths or misestimates the path count by two or more,

\begin{equation}\tag{11}\label{eq:11}
\fitmath{\mathcal F_{\mathrm{sev}}=
\{L>0\ \text{and}\ R<0.5\}\ \cup\ 
\{|\widehat L-L|\ge 2\}.}
\end{equation}

In noise-only trials, the empirical false-alarm probability is \(P_{\mathrm{FA,emp}}=\Pr(\widehat L>0\mid L=0)\). Note that path recall \(R\), position-matched detection \(D_{\mathrm{pos}}\), amplitude-consistent detection \(D_{\mathrm{amp}}\), and severe failure \(\mathcal F_{\mathrm{sev}}\) capture distinct operational properties and should not be conflated.

\section{Continuous SAGE--Brent Refinement}\label{sec:3}

\subsection{Conditional Objective and Alternating Continuous Updates}\label{sec:3.1}

During outer iteration \(i\), we isolate path \(\ell\) by subtracting the latest reconstructions of all other active paths from the observation \(\mathbf y\). This subtraction yields the SAGE conditional observation (hidden-data estimate)

\begin{equation}\tag{12}\label{eq:12}
\fitmath{\mathbf z_\ell^{(i)}=\mathbf y-
\sum_{j<\ell}\widehat\alpha_j^{(i+1)}
\mathbf a(\widehat{\boldsymbol\theta}_j^{(i+1)})-
\sum_{j>\ell}\widehat\alpha_j^{(i)}
\mathbf a(\widehat{\boldsymbol\theta}_j^{(i)}).}
\end{equation}

For any candidate parameter pair \(\boldsymbol\theta=(\tau,\epsilon)^{\mathsf T}\), the least-squares complex gain estimate has the closed form

\begin{equation}\tag{13}\label{eq:13}
\fitmath{\widehat\alpha(\boldsymbol\theta)=
\frac{\mathbf a(\boldsymbol\theta)^{\mathsf H}\mathbf z_\ell}
{\|\mathbf a(\boldsymbol\theta)\|_2^2}.}
\end{equation}

Substituting~\eqref{eq:13} into the conditional residual sum of squares concentrates the complex gain out of the optimization. Minimizing the residual energy then reduces to maximizing the single-path matched-energy objective

\begin{equation}\tag{14}\label{eq:14}
\fitmath{J_\ell(\tau,\epsilon)=
\frac{|\mathbf a(\tau,\epsilon)^{\mathsf H}\mathbf z_\ell|^2}
{\|\mathbf a(\tau,\epsilon)\|_2^2}.}
\end{equation}

Following the SAGE principle \cite{FesslerHero1994SAGE}, each M-step updates a single path by maximizing~\eqref{eq:14} over continuous delay and Doppler scale within local bounds set by the ambiguity main lobe and adjacent paths.

Evaluating~\eqref{eq:14} across a dense two-dimensional grid over \((\tau,\epsilon)\) scales with the product of the two axis grid sizes. To avoid this quadratic growth, we initialize at a coarse-grid peak \((\tau^{(0)},\epsilon^{(0)})\) and alternate one-dimensional continuous searches along delay and Doppler scale,

\begin{equation}\tag{15}\label{eq:15}
\fitmath{\tau^{(q+1)}=\arg\max_{\tau\in\mathcal I_\tau}
J_\ell(\tau,\epsilon^{(q)}),}
\end{equation}

\begin{equation}\tag{16}\label{eq:16}
\fitmath{\epsilon^{(q+1)}=\arg\max_{\epsilon\in\mathcal I_\epsilon}
J_\ell(\tau^{(q+1)},\epsilon).}
\end{equation}

We solve each one-dimensional maximization using Brent's method \cite{Brent1973Minimization}, which combines golden-section bracketing with successive parabolic interpolation. Whenever a parabolic step falls outside the bracket or fails to improve the objective fast enough, the solver reverts to a golden-section step. Furthermore, we accept an updated coordinate only if it does not decrease \(J_\ell(\tau,\epsilon)\); otherwise, we retain the previous coordinate. This acceptance rule guarantees monotonic objective values across alternating steps.

\begin{figure*}[!t]
\centering
\includegraphics[width=\textwidth,height=.72\textheight,keepaspectratio]{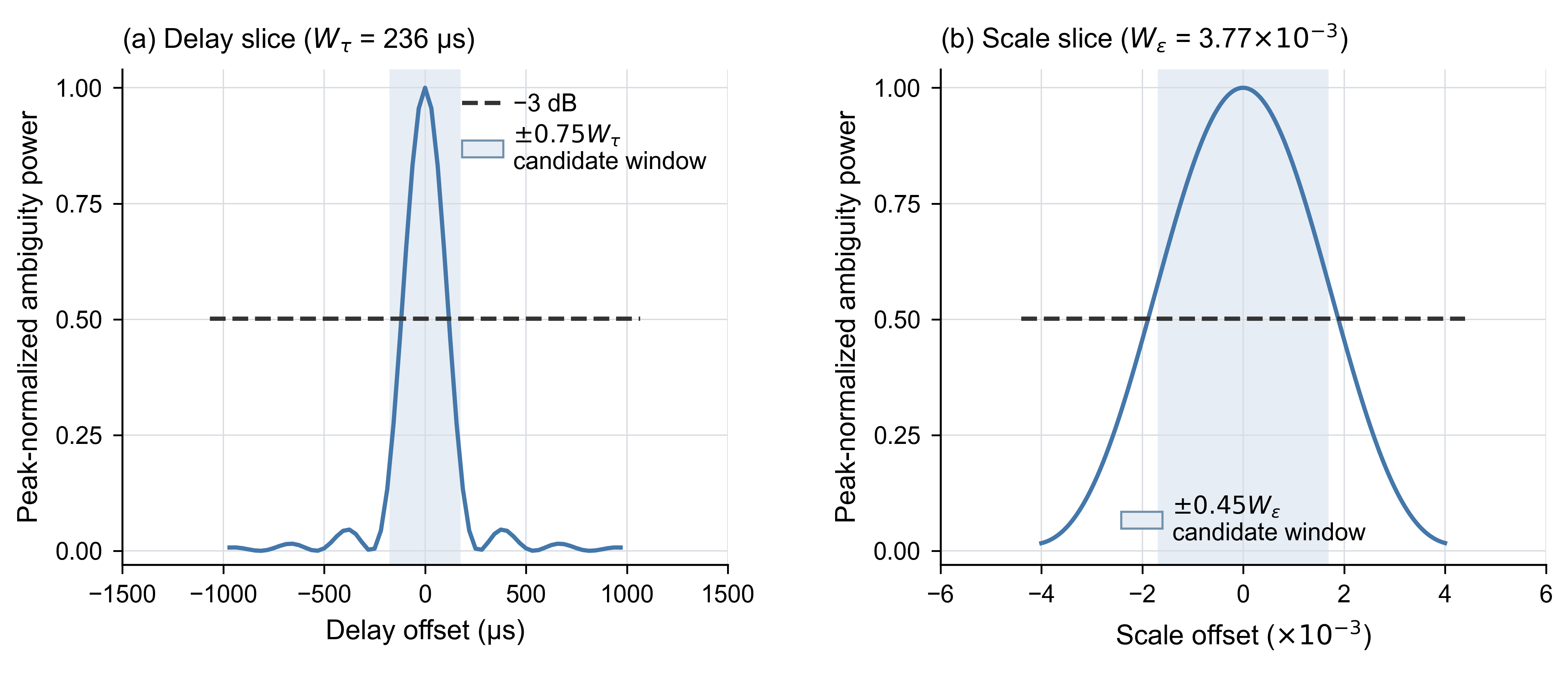}
\caption{LFM ambiguity function and candidate search windows: (a) delay slice; (b) scale slice.}\label{fig:2}

\end{figure*}

Fig.~\ref{fig:2} illustrates the peak-normalized ambiguity-power slices of a 25 ms LFM probe computed via~\eqref{eq:5a}. The horizontal dashed lines mark the \ensuremath{-}3 dB calibration level, and the shaded bands indicate the default local search windows of \(\pm0.75W_\tau\) in delay and \(\pm0.45W_\epsilon\) in Doppler scale when no adjacent paths are present. During cyclic multipath refinement, if another active path lies on either side along the delay axis, we place the interval boundary at the midpoint between the two path delays; if no neighbor lies on that side, we retain the main-lobe-width bound. Along the Doppler-scale axis, the search always uses a local window centered on the current scale estimate.

\subsection{Search Intervals and Refinement Procedure}\label{sec:3.2}

Let \((\tau_c,\epsilon_c)\) denote the current parameter estimate, and let \(\mathcal B_\tau=[0,\tau_{\max}]\) and \(\mathcal B_\epsilon=[\epsilon_{\min},\epsilon_{\max}]\) denote the physically feasible parameter ranges. During initial candidate validation and residual redetection, single-path refinement searches over the ambiguity-constrained intervals

\begin{equation}\tag{17}\label{eq:17}
\fitmath{\begin{gathered}
\mathcal I_{\tau,\mathrm{AF}}=
[\tau_c-m_\tau W_\tau,\ \tau_c+m_\tau W_\tau]
\cap\mathcal B_\tau,\\
\mathcal I_{\epsilon,\mathrm{AF}}=
[\epsilon_c-m_\epsilon W_\epsilon,\ \epsilon_c+m_\epsilon W_\epsilon]
\cap\mathcal B_\epsilon,
\end{gathered}}
\end{equation}

with \(m_\tau=0.75\) and \(m_\epsilon=0.45\). In two-dimensional candidate refinement, we recenter both windows on the latest coordinates at each alternating pass. In one-dimensional delay-only refinement, the search remains within the initial delay bracket.

During cyclic SAGE refinement across multiple active paths, we first sort the active paths in ascending order of their current delay estimates. For path \(\ell\) with immediate left and right delay neighbors \(\tau_-\) and \(\tau_+\), we construct the delay search bracket \([b_-,b_+]\), where

\begin{equation}\tag{17a}\label{eq:17a}
\fitmath{\begin{gathered}
b_-=\begin{cases}
(\tau_-+\tau_\ell)/2,&\text{if left neighbor exists},\\
\max(0,\tau_\ell-m_\tau W_\tau),&\text{otherwise},
\end{cases}\\
b_+=\begin{cases}
(\tau_\ell+\tau_+)/2,&\text{if right neighbor exists},\\
\min(\tau_{\max},\tau_\ell+m_\tau W_\tau),&\text{otherwise}.
\end{cases}
\end{gathered}}
\end{equation}

We clip both endpoints to \(\mathcal B_\tau\) while ensuring that the bracket contains \(\tau_\ell\). When a neighbor exists on a given side, the midpoint bound replaces the main-lobe bound in~\eqref{eq:17} so that adjacent paths do not cross or collapse onto the same peak. Along the Doppler-scale axis, we always apply the local window \(\mathcal I_{\epsilon,\mathrm{AF}}\) in~\eqref{eq:17} without midpoint partitioning. If a delay bracket ever degenerates to zero width, the delay search reverts to \(\mathcal I_{\tau,\mathrm{AF}}\) and the scale search reverts to \(\mathcal B_\epsilon\). We apply this delay-midpoint rule even when paths have close delays and distinct Doppler scales; Section~\ref{sec:8.2} discusses the resulting subresolution limits.

Within each single-path update, the alternating Brent search runs for at most four inner passes. It stops early if the relative changes in both \(\tau\) and \(\epsilon\) fall below \(10^{-3}\), or if the absolute changes fall below \(10^{-4}/f_s\) in delay and \(10^{-4}W_\epsilon\) in scale. Known-order two-dimensional estimation executes up to 10 outer SAGE cycles, whereas unknown-order RC-SAGE--Brent executes up to 12 closed-loop outer cycles. Each one-dimensional Brent call allows at most 200 objective evaluations.

\begin{algorithm}[!t]
\caption{Continuous SAGE--Brent single-path refinement}\label{alg:1}
\footnotesize\begin{algorithmic}[1]

\REQUIRE Conditional observation \(\mathbf z_\ell\), current parameters \((\tau_c,\epsilon_c)\), active path set, physical bounds \(\mathcal B_\tau,\mathcal B_\epsilon\), and main-lobe widths \(W_\tau,W_\epsilon\).

\STATE Set local search brackets: use~\eqref{eq:17} during candidate refinement, or use~\eqref{eq:17a} for delay and~\eqref{eq:17} for Doppler scale during cyclic multipath refinement. Clamp \((\tau_c,\epsilon_c)\) to these brackets. Allow up to four inner alternating passes during candidate refinement and one delay--scale pass per path per outer cycle during unknown-order cyclic refinement.

\STATE Fix \(\epsilon\) and maximize \(J_\ell(\tau,\epsilon)\) in~\eqref{eq:14} over \(\mathcal I_\tau\) via Brent's method; update \(\tau\) only if the objective does not decrease.

\STATE Fix the updated \(\tau\) and maximize \(J_\ell(\tau,\epsilon)\) in~\eqref{eq:14} over \(\mathcal I_\epsilon\) via Brent's method; update \(\epsilon\) only if the objective does not decrease.

\STATE Update the complex gain \(\widehat\alpha_\ell\) via~\eqref{eq:13}. If the stopping tolerances or inner-pass limit are reached, proceed to Step 5; otherwise, return to Step 2.

\STATE Return the refined parameters \((\widehat\alpha_\ell,\widehat\tau_\ell,\widehat\epsilon_\ell)\) and the number of objective evaluations.

\end{algorithmic}

\end{algorithm}

Under known model order, the estimator applies Algorithm~\ref{alg:1} sequentially to each path within the outer SAGE loop of~\eqref{eq:12}. Under unknown model order, the closed-loop estimator invokes Algorithm~\ref{alg:1} to refine newly detected candidates prior to validation, to update surviving paths during cyclic refinement, and to refine residual peaks during redetection.

\section{Closed-Loop Unknown-Order Estimation with Residual Feedback}\label{sec:4}

RC-SAGE--Brent jointly tracks the active path set and the observation residual generated by its current reconstruction. Let \(\widehat{\mathcal S}^{(t)}\) and \(\mathbf r^{(t)}\) denote the active path set and observation residual at outer cycle \(t\), respectively. Each cycle executes continuous refinement, conditional gain refitting, candidate validation, resolution pruning, and full-domain residual redetection to produce \((\widehat{\mathcal S}^{(t+1)},\mathbf r^{(t+1)})\).

The observation residual acts as the feedback state for the next cycle. Continuous refinement suppresses strong-path cancellation residuals, pruning eliminates invalid or duplicate components, and redetection recovers weak paths that were previously masked. Any structural modification to the path set triggers another round of continuous refinement and cancellation. Standalone thresholding, backward deletion, or local parameter refinement cannot achieve this bidirectional correction in isolation.

\subsection{Path Discovery, Validation, and Reestimation}\label{sec:4.1}

Given the current residual \(\mathbf r\), we first evaluate the normalized matched-filter energy across the coarse candidate grid \(\mathcal G\),

\begin{equation}\tag{18}\label{eq:18}
\fitmath{T(\boldsymbol\theta_g;\mathbf r)=
\frac{|\mathbf a(\boldsymbol\theta_g)^{\mathsf H}\mathbf r|^2}
{\|\mathbf a(\boldsymbol\theta_g)\|_2^2}.}
\end{equation}

Exploiting the relationship between the median and variance of exponentially distributed sample energies under complex Gaussian noise, we estimate the residual noise variance robustly via

\begin{equation}\tag{19}\label{eq:19}
\fitmath{\widehat\sigma_r^2=
\frac{\operatorname{median}_n |r_n|^2}{\ln 2}.}
\end{equation}

For \(M_{\mathrm{eff}}\) effectively independent search cells, we set the CFAR candidate detection threshold as

\begin{equation}\tag{20}\label{eq:20}
\fitmath{\gamma_r=\widehat\sigma_r^2
\ln\!\left(\frac{M_{\mathrm{eff}}}{P_{\mathrm{FA}}}\right),}
\end{equation}

with nominal false-alarm probability \(P_{\mathrm{FA}}=10^{-3}\) and maximum path count \(L_{\max}=10\). The effective cell count \(M_{\mathrm{eff}}\) accounts for correlation across adjacent candidate waveforms, and our noise-only trials verify the resulting empirical false-alarm rate.

We refine each coarse correlation peak using Algorithm~\ref{alg:1}. The estimator admits a refined candidate if and only if its matched energy exceeds \(\gamma_r\) and its normalized distance from all currently retained paths exceeds \(d_{\min}\). We set \(d_{\min}=0.8W_\tau\) in one-dimensional delay estimation and \(d_{\min}=0.6\) (in normalized delay--scale units) in two-dimensional estimation.

For a fixed active parameter set \(\widehat{\boldsymbol\Theta}\), the joint least-squares gain vector is

\begin{equation}\tag{21}\label{eq:21}
\fitmath{\widehat{\boldsymbol\alpha}=
\arg\min_{\boldsymbol\alpha}
\|\mathbf y-\mathbf A(\widehat{\boldsymbol\Theta})
\boldsymbol\alpha\|_2^2
=\mathbf A(\widehat{\boldsymbol\Theta})^\dagger\mathbf y.}
\end{equation}

Inside the loop, we update path gains through cyclic pathwise conditional least-squares steps. Specifically, for each active path, we add its current waveform reconstruction back to the residual \(\mathbf r\) to form its conditional observation, update its complex gain via~\eqref{eq:13}, and subtract the updated reconstruction before moving to the next path. When delays and Doppler scales are held fixed, this cyclic coordinate descent converges toward the normal equations of~\eqref{eq:21}. Whenever two active paths violate the minimum separation \(d_{\min}\), resolution pruning retains only the path with the larger individual matched energy and recomputes the residual from the surviving set. Thus, validation blocks noise spikes, resolution pruning removes duplicate detections, and cyclic conditional updates reduce order-dependent sequential bias.

\subsection{Residual Redetection and Closed-Loop Procedure}\label{sec:4.2}

After each pass of parameter refinement, pruning, and gain updates, we recompute the observation residual from the active path set,

\begin{equation}\tag{22}\label{eq:22}
\fitmath{\mathbf r=\mathbf y-
\mathbf A(\widehat{\boldsymbol\Theta})
\widehat{\boldsymbol\alpha}.}
\end{equation}

Following initial candidate validation, the estimator performs an initial residual redetection pass to capture weak paths exposed by strong-path refinement. While the active path count remains below \(L_{\max}\), the algorithm scans the full coarse grid over \(\mathbf r\), refines the strongest peak via Algorithm~\ref{alg:1}, and admits the refined candidate if it satisfies both the energy threshold \(\gamma_r\) and the minimum spacing \(d_{\min}\). This initial redetection stage ends as soon as the strongest coarse or refined peak fails either check or the path count reaches \(L_{\max}\). The estimator then enters the outer refinement--pruning--redetection loop. During each outer cycle with a nonempty path set, the algorithm tests at most one dominant coarse-grid residual peak and admits it if valid; if the active path set is empty, it skips redetection. Standard unknown-order RC-SAGE--Brent never replaces existing paths once the path count reaches \(L_{\max}\).

Even when redetection admits no new candidate or the active path count sits at \(L_{\max}\), cyclic refinement and pruning of the existing paths continue. The outer loop terminates when all normalized parameter updates fall below \(10^{-3}\) and the path set experiences no additions or deletions, or when 12 outer cycles complete. Because this stopping rule depends on parameter convergence and support stability rather than a single-pass energy threshold, weak paths initially masked by strong-path cancellation residuals can still emerge after subsequent refinement and pruning cycles.

\begin{figure}[!t]
\centering
\includegraphics[width=\columnwidth,height=.66\textheight,keepaspectratio]{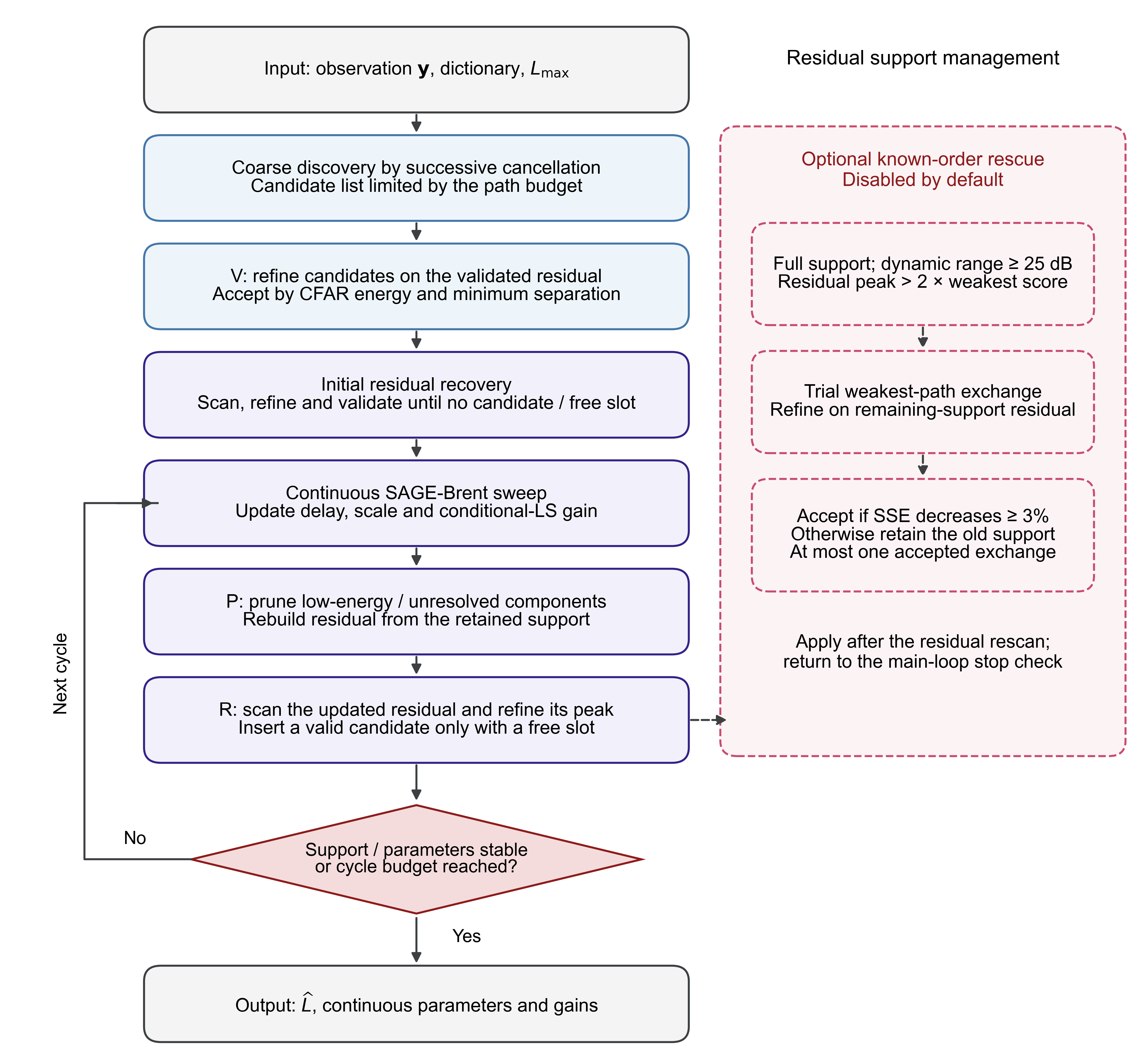}
\caption{Residual closed loop for continuous refinement and unknown-order path selection and updating.}\label{fig:3}

\end{figure}

Fig.~\ref{fig:3} illustrates the interaction between continuous parameter refinement and closed-loop support management. The SAGE--Brent inner module updates continuous delays, Doppler scales, and gains, while the outer residual feedback loop governs candidate validation, pruning, redetection, and termination. The standard workflow proceeds through coarse candidate discovery, sorted validation, initial redetection, and cyclic refinement--pruning--redetection. The dashed block on the right indicates an optional known-order replacement test that remains disabled in standard unknown-order estimation. That optional step activates only in known-order diagnostics when the path count reaches \(L_{\max}\) and the estimated gain dynamic range exceeds 25 dB.

\begin{algorithm}[!t]
\caption{RC-SAGE--Brent unknown-order estimation}\label{alg:2}
\footnotesize\begin{algorithmic}[1]

\REQUIRE Observation \(\mathbf y\), coarse grid \(\mathcal G\), false-alarm parameter \(P_{\mathrm{FA}}\), minimum separation \(d_{\min}\), maximum order \(L_{\max}\), and continuous-search bounds.

\STATE Initialize \(\widehat{\mathcal S}\leftarrow\varnothing\). Run coarse-grid successive cancellation on \(\mathbf y\) up to \(L_{\max}\) candidates without early CFAR stopping, and sort the candidates by descending preliminary gain magnitude.

\STATE Validate candidates sequentially. Reconstruct \(\mathbf r\) from the current set \(\widehat{\mathcal S}\), compute \(\widehat\sigma_r^2\) and \(\gamma_r\) via~\eqref{eq:19} and~\eqref{eq:20}, and refine the candidate over~\eqref{eq:17}. Admit the candidate only if its refined energy satisfies \(J_{\mathrm{own}}\ge\gamma_r\), its gain is nonzero, and its distance from all retained paths exceeds \(d_{\min}\). Continue testing remaining candidates after any rejection.

\STATE Initial residual redetection: while \(|\widehat{\mathcal S}|<L_{\max}\), update \(\mathbf r\) and \(\gamma_r\), and find the strongest coarse-grid peak across \(\mathcal G\). Terminate this stage if the coarse peak falls below \(\gamma_r\) or violates \(d_{\min}\); otherwise, refine it via Algorithm~\ref{alg:1}. Terminate this stage if the refined peak fails \(\gamma_r\) or \(d_{\min}\); otherwise, admit it into \(\widehat{\mathcal S}\) and repeat Step 3.

\STATE Start an outer cycle. For each path in \(\widehat{\mathcal S}\), add its current reconstruction back to \(\mathbf r\) to form the conditional observation, update its delay over~\eqref{eq:17a}, its Doppler scale over~\eqref{eq:17}, and its gain via~\eqref{eq:13}, and subtract the updated reconstruction. Record whether any normalized coordinate change exceeds \(10^{-3}\).

\STATE Recompute \(\gamma_r\) from the end-of-cycle residual. Screen paths in descending order of conditional matched energy \(J_{\mathrm{own}}\), retaining only those that exceed \(\gamma_r\) and satisfy \(d_{\min}\) relative to already retained paths. Record any deletions and rebuild \(\mathbf r\) from the surviving paths.

\STATE If \(\widehat{\mathcal S}\neq\varnothing\) and \(|\widehat{\mathcal S}|<L_{\max}\), find the strongest coarse-grid peak of \(\mathbf r\) across \(\mathcal G\). If that peak exceeds the Step 5 threshold \(\gamma_r\) and satisfies \(d_{\min}\), refine it via Algorithm~\ref{alg:1} and recheck both criteria. If valid, admit the path, rebuild \(\mathbf r\), and record the addition. Do not replace paths when \(|\widehat{\mathcal S}|=L_{\max}\).

\STATE If all coordinate updates fall within \(10^{-3}\) and no paths were added or deleted during the cycle, or if 12 outer cycles have completed, proceed to Step 8; otherwise, return to Step 4.

\STATE Apply backward deletion and joint least-squares refitting to \(\widehat{\mathcal S}\). Evaluate the model-order score~\eqref{eq:22a} using \(\lambda=1\) for PN probes or the residual-adaptive penalty~\eqref{eq:22b}--\eqref{eq:22c} for LFM probes. Return the order \(\widehat L\) that minimizes \(\mathcal C(k)\), along with the corresponding continuous parameters and complex gains.

\end{algorithmic}

\end{algorithm}

Algorithm~\ref{alg:2} summarizes the standard two-dimensional unknown-order estimator. The one-dimensional delay-only variant fixes \(\epsilon=0\) and orders path updates by their cancellation impact. During initial redetection, the one-dimensional variant checks minimum spacing on the coarse grid prior to refinement, whereas the two-dimensional variant checks spacing both before and after refinement. Both versions share the same four-stage architecture—coarse discovery, sorted validation, initial redetection, and cyclic refinement--pruning--redetection—and neither allows path replacement at \(L_{\max}\). To maintain consistent energy scaling across dictionary resolutions, we scale coarse-grid peak energies and detection thresholds by the search upsampling factor.

\subsection{Residual-Adaptive Model-Order Selection}\label{sec:4.3}

Given the \(L_c\le L_{\max}\) candidate paths returned by the closed loop, we apply greedy backward deletion to construct nested path subsets of orders \(k\in\{0,1,\dots,L_c\}\). Refitting each subset via joint least squares yields the residual sum of squares \(\operatorname{RSS}_k\). We then select the final model order \(\widehat L\) by minimizing the penalized likelihood criterion

\begin{equation}\tag{22a}\label{eq:22a}
\fitmath{\mathcal C(k)=2N\log\frac{\operatorname{RSS}_k}{N}
+\lambda\left[(4k+1)\log(2N)+2k\log M\right],}
\end{equation}

where \(M=250\) denotes the reference dictionary size. For pseudorandom noise (PN) probes, we use the fixed weight \(\lambda=1\). For LFM probes, however, strong delay--scale coupling along the ambiguity ridge causes structured high-SNR cancellation residuals to mimic additional weak arrivals. A constant penalty weight \(\lambda\) cannot preserve genuine weak paths at low SNR while suppressing ridge-induced false paths at high SNR. To resolve this tradeoff without knowing the true SNR, we compute an observable signal-to-residual energy ratio \(q\) directly from the full-candidate fit,

\begin{equation}\tag{22b}\label{eq:22b}
\fitmath{q=10\log_{10}
\frac{\max\{\|\mathbf y\|_2^2-\operatorname{RSS}_{L_c},\varepsilon\}}
{\max\{\operatorname{RSS}_{L_c},\varepsilon\}},}
\end{equation}

where \(\varepsilon>0\) prevents division by zero. Because \(q\) compares the fitted signal energy against the unmodeled residual energy, it increases monotonically with effective SNR. Using a disjoint development set of 100 LFM scenarios, we calibrated and froze the piecewise-linear penalty schedule

\begin{equation}\tag{22c}\label{eq:22c}
\fitmath{\lambda(q)=\operatorname{clip}\{2+1.5(q-10),1,20\}.}
\end{equation}

The development set spanned SNRs of 10, 15, 20, and 25 dB across five stress regimes and true orders \(L\in\{1,\dots,5\}\). We selected the slope and intercept in~\eqref{eq:22c} to maximize order accuracy subject to an underestimation rate below 5\%, and then evaluated the frozen rule on an independent 100-scenario validation set. Because \(\|\mathbf y\|_2^2\) and \(\operatorname{RSS}_{L_c}\) are already available from the joint least-squares evaluations in~\eqref{eq:22a}, computing~\eqref{eq:22b} and~\eqref{eq:22c} adds only scalar arithmetic and requires no additional matrix factorizations.

\subsection{Estimators and Parameter Baseline}\label{sec:4.4}

Throughout the experiments, we compare two configurations of our method to separate continuous parameter refinement from closed-loop support management. ``SAGE--Brent'' denotes the standalone continuous refinement module (Algorithm~\ref{alg:1}), which performs cyclic SAGE updates and alternating Brent searches under a fixed, known path count. ``RC-SAGE--Brent'' denotes the complete closed-loop estimator (Algorithm~\ref{alg:2}), which integrates SAGE--Brent refinement with candidate validation, resolution pruning, residual redetection, and model-order selection. Table~\ref{tab:1} summarizes the baseline algorithmic parameters used across all synthetic benchmarks.

\begin{table}[!t]
\caption{Baseline RC-SAGE--Brent parameters}\label{tab:1}
\centering\footnotesize
\setlength{\tabcolsep}{3pt}
\begin{tabularx}{\columnwidth}{YYY}
\toprule

Parameter & Description & Value / Rule \\

\midrule

\(P_{\mathrm{FA}}\) & Nominal false-alarm parameter for CFAR threshold & \(10^{-3}\) \\

\(\widehat\sigma_r^2\) & Robust residual noise variance estimator & \(\operatorname{median}|r_n|^2/\ln2\) \\

\(L_{\max}\) & Maximum allowed path count & 10 \\

\(d_{\min}\) & Minimum inter-path resolution gate & \(0.8W_\tau\) (1-D delay); \(0.6\) (2-D normalized) \\

Redetection & Full-domain coarse rescan after gain updates & Enabled \\

Initial redetection stop & Strongest peak below threshold or \(L_{\max}\) reached & Enter outer refinement loop \\

Outer loop termination & Coordinate changes \(\le 10^{-3}\) with stable support, or 12 cycles & Structural stability or cycle cap \\

\bottomrule

\end{tabularx}

\end{table}

\section{Theoretical Analysis of Cancellation Residuals and Weak-Path Detection}\label{sec:5}

\subsection{Cancellation-Residual Energy and Local Curvature}\label{sec:5.1}

Consider a noiseless single-path observation \(\mathbf x=\alpha\mathbf a_0\) with true parameter vector \(\boldsymbol\theta_0\), fitted by a mismatched candidate waveform \(\mathbf a_\delta=\mathbf a(\boldsymbol\theta_0+\boldsymbol\delta)\) of equal norm. When the complex gain is estimated optimally via~\eqref{eq:13}, subtracting the reconstructed path leaves the orthogonal projection residual

\begin{equation}\tag{23}\label{eq:23}
\fitmath{\mathbf e_\delta=\alpha
\left(\mathbf I-
\frac{\mathbf a_\delta\mathbf a_\delta^{\mathsf H}}
{\|\mathbf a_\delta\|_2^2}\right)\mathbf a_0.}
\end{equation}

The ratio of cancellation-residual energy to true path energy is therefore

\begin{equation}\tag{24}\label{eq:24}
\fitmath{\frac{\|\mathbf e_\delta\|_2^2}{\|\mathbf x\|_2^2}
=1-|\rho(\boldsymbol\theta_0,
\boldsymbol\theta_0+\boldsymbol\delta)|^2.}
\end{equation}

Equation~\eqref{eq:24} shows that even with an optimal complex gain, any off-grid parameter offset \(\boldsymbol\delta\) yielding \(|\rho(\boldsymbol\theta_0,\boldsymbol\theta_0+\boldsymbol\delta)|<1\) prevents complete cancellation. The unmodeled energy remains in the observation residual as a structured waveform dictated by the probe ambiguity function.

Let \(\mathbf D=[\partial\mathbf a/\partial\tau, \partial\mathbf a/\partial\epsilon]_{\boldsymbol\theta_0}\) denote the Jacobian matrix at \(\boldsymbol\theta_0\), let \(E_a=\|\mathbf a_0\|_2^2\) denote the waveform energy, and let \(\mathbf P_\perp=\mathbf I-\mathbf a_0\mathbf a_0^{\mathsf H}/E_a\) denote the projector onto the orthogonal complement of \(\mathbf a_0\). Assuming that \(\mathbf a(\boldsymbol\theta)\) is twice continuously differentiable and the parameter offset \(\boldsymbol\delta\) is small, a second-order Taylor expansion of~\eqref{eq:24} yields

\begin{equation}\tag{25}\label{eq:25}
\fitmath{\begin{gathered}
1-|\rho|^2=
\boldsymbol\delta^{\mathsf T}\mathbf Q\boldsymbol\delta
+\mathcal O(\|\boldsymbol\delta\|_2^3),\\
\mathbf Q=\frac{1}{E_a}
\operatorname{Re}\{\mathbf D^{\mathsf H}
\mathbf P_\perp\mathbf D\}.
\end{gathered}}
\end{equation}

Under additive white complex Gaussian noise \(\mathcal{CN}(\mathbf0,\sigma^2\mathbf I)\) with unknown complex gain \(\alpha\), the equivalent Fisher information matrix for \((\tau,\epsilon)\)—obtained as the Schur complement of the gain block—is \cite{Kay1993Estimation}

\begin{equation}\tag{26}\label{eq:26}
\fitmath{\mathbf J_{\tau,\epsilon\mid\alpha}=
\frac{2|\alpha|^2}{\sigma^2}
\operatorname{Re}\{\mathbf D^{\mathsf H}
\mathbf P_\perp\mathbf D\}
=\frac{2|\alpha|^2E_a}{\sigma^2}\mathbf Q.}
\end{equation}

Equation~\eqref{eq:26} establishes that the local curvature matrix \(\mathbf Q\) of the cancellation residual is directly proportional to the conditional Fisher information matrix \(\mathbf J_{\tau,\epsilon\mid\alpha}\). When \(\mathbf Q\) is poorly conditioned, as occurs along the delay--scale coupling ridge of an LFM waveform, the partial derivative vectors in \(\mathbf D\) become nearly collinear. In that regime, parameter estimation variance increases along the ridge, while coupled off-grid errors transverse to the ridge leave substantial residual energy. Note that~\eqref{eq:25} and~\eqref{eq:26} describe local behavior within the main lobe of resolvable paths; characterizing far-field sidelobes and closely spaced multipath interactions requires global multimodal analysis.

\subsection{Residual Strong-Path Interference and Weak-Path Detectability}\label{sec:5.2}

Consider a two-path near--far scenario in which a strong path \(\mathbf x_s=\alpha_s\mathbf a_s\) leaves a cancellation residual \(\mathbf e_s\) due to parameter mismatch, and let \(\mathbf b_w=\mathbf a_w/\|\mathbf a_w\|_2\) denote the unit-norm waveform of a weak path with complex gain \(\alpha_w\). By the Cauchy--Schwarz inequality, the interference energy that \(\mathbf e_s\) induces at the weak-path matched filter satisfies

\begin{equation}\tag{27}\label{eq:27}
\fitmath{|\mathbf b_w^{\mathsf H}\mathbf e_s|^2
\le \|\mathbf e_s\|_2^2
=|\alpha_s|^2E_a(1-|\rho_s|^2),}
\end{equation}

where \(\rho_s\) is the ambiguity correlation between the true and estimated strong-path parameters. Because the noiseless weak path produces a matched-filter amplitude of \(|\alpha_w|\sqrt{E_a}\), a conservative worst-case sufficient condition for the weak path to exceed the detection threshold \(\sqrt{\gamma_r}\) with noise margin \(\eta\sigma\) is

\begin{equation}\tag{28}\label{eq:28}
\fitmath{|\alpha_w|\sqrt{E_a}
-|\alpha_s|\sqrt{E_a(1-|\rho_s|^2)}
>\sqrt{\gamma_r}+\eta\sigma .}
\end{equation}

Equation~\eqref{eq:28} directly links the near--far ratio \(\Gamma_{\mathrm{NF}}=20\log_{10}(|\alpha_s|/|\alpha_w|)\), the strong-path correlation \(|\rho_s|\), and the CFAR threshold \(\gamma_r\). As \(\Gamma_{\mathrm{NF}}\) grows, the maximum tolerable mismatch \(1-|\rho_s|^2\) shrinks inversely with \(|\alpha_s/\alpha_w|^2\). Continuous SAGE--Brent refinement drives \(|\rho_s|\) toward unity, which lowers the cancellation-residual norm \(\|\mathbf e_s\|_2\) and tightens the interference bound in~\eqref{eq:27}. Subsequent full-domain residual redetection can then identify weak paths that were initially buried under \(\mathbf e_s\). We verify this mechanism empirically through the mismatch curves in Figs.~\ref{fig:1}(b) and (c), the near--far diagnostics in Fig.~\ref{fig:7}, and the factorial ablations in Section~\ref{sec:7.2}.

\subsection{Update Properties and Complexity}\label{sec:5.3}

Within Algorithm~\ref{alg:1}, each one-dimensional Brent step updates a coordinate only if doing so does not decrease \(J_\ell(\tau,\epsilon)\) in~\eqref{eq:14}, and~\eqref{eq:13} computes the exact least-squares complex gain for the updated coordinates. Consequently, every delay, scale, and gain update monotonically nondecreases the single-path matched energy \(J_\ell\) for a fixed conditional observation \(\mathbf z_\ell\). Because \(J_\ell\) is bounded above by the finite energy \(\|\mathbf z_\ell\|_2^2\), the sequence of objective values for a fixed conditional observation converges. In the full closed-loop estimator (Algorithm~\ref{alg:2}), however, candidate additions and pruning steps alter the active support and redefine the conditional observations across outer cycles. Therefore, single-path monotonicity does not imply global monotonic convergence of the outer loop; instead, Algorithm~\ref{alg:2} guarantees termination through its coordinate-stability check and its hard cap of 12 outer cycles.

To compare search complexity, let \(N_\tau\) and \(N_\epsilon\) denote the numbers of fine-grid points along the delay and Doppler-scale axes. Exhaustively evaluating a two-dimensional fine grid requires \(\mathcal O(N_\tau N_\epsilon)\) objective evaluations per path update. By contrast, alternating one-dimensional Brent searches require \(q_\tau\) and \(q_\epsilon\) evaluations along the two axes per inner pass. Over \(L\) active paths, \(K_{\mathrm{in}}\) inner passes, and \(K_{\mathrm{out}}\) outer SAGE cycles, the total number of scalar objective evaluations scales as

\begin{equation}\tag{29}\label{eq:29}
\fitmath{\mathcal O(K_{\mathrm{out}} L K_{\mathrm{in}}(q_\tau+q_\epsilon)).}
\end{equation}

Note that~\eqref{eq:29} tallies scalar evaluations of \(J_\ell(\tau,\epsilon)\) rather than total wall-clock time, which also includes waveform synthesis, least-squares projections, and memory management. We therefore report measured single-thread execution times and resident set size (RSS) increments separately in Section~\ref{sec:7.5}.

During final model-order selection, both the fixed and residual-adaptive criteria perform backward deletion and joint least-squares refitting over at most \(L_c\) candidates. Because the adaptive penalty in~\eqref{eq:22b} and~\eqref{eq:22c} reuses \(\|\mathbf y\|_2^2\) and \(\operatorname{RSS}_{L_c}\) from those fits, it introduces only \(\mathcal O(1)\) scalar operations and leaves asymptotic time and memory complexity unchanged. In single-thread paired timing across 100 identical candidate sets, the fixed and adaptive rules required 0.26155 s and 0.26171 s on average—a negligible difference of 0.064\%.

Empirical evaluation counts align closely with~\eqref{eq:29}. For a single LFM path at 20 dB SNR, one-dimensional SAGE--Brent achieves a delay root-mean-squared error (RMSE) of 0.220 \(\mu\mathrm{s}\) using 981 objective evaluations, whereas a one-dimensional fine grid requires 19201 evaluations and still stalls at 0.713 \(\mu\mathrm{s}\). For a single two-dimensional PN path at 20 dB SNR, alternating SAGE--Brent reaches a delay RMSE of 0.189 \(\mu\mathrm{s}\) in 4909 evaluations, compared with \(1.56\times10^6\) evaluations for a full two-dimensional fine grid (Fig.~\ref{fig:5}). Appendices~\ref{app:A}--\ref{app:C} provide the complete mathematical derivations.

\section{Experimental Settings and Statistical Methods}\label{sec:6}

\subsection{Experimental Design and Statistical Methods}\label{sec:6.1}

Our experimental evaluation spans six benchmark suites: single-path parameter accuracy, known-order multipath reconstruction, unknown-order path-set estimation, two-dimensional delay--scale estimation, measured impulse-response replay, and single-thread computational complexity. Within each synthetic scenario, every competing estimator receives the exact same noiseless channel realization, additive noise realization, and probe waveform, enabling direct scenario-wise paired comparisons.

The one-dimensional unknown-order benchmark tests true path counts \(L\in\{0,1,2,3,5,8\}\) across SNRs of 0, 10, and 20 dB. At each SNR, we generate 80, 80, 80, 60, 40, and 20 independent scenarios for the six respective orders, yielding 360 scenarios per SNR and 1080 scenarios overall. Path delays, complex phases, and noise realizations are drawn independently from prescribed distributions. The 240 noise-only scenarios (\(L=0\)) quantify empirical false-alarm rates. The two-dimensional unknown-order benchmark replicates this exact design of 1080 scenarios while additionally drawing a continuous Doppler scale \(\epsilon_\ell\) for each path. Allocating more trials to lower orders sharpens statistical precision where path-count transitions are frequent, while fewer trials at \(L=8\) keep multi-method two-dimensional runtimes tractable.

The primary two-dimensional known-order benchmark comprises 280 scenarios: 80 single-path runs covering PN and LFM probes at 10 and 20 dB SNR; 120 two-path near--far runs spanning NFRs of 15, 25, and 35 dB across two Doppler-scale separations; and 80 random-multipath runs with \(L\in\{3,5\}\) at 10 and 20 dB SNR. An additional 160 paired random-multipath scenarios sweep SNR across 0, 10, 20, and 30 dB for \(L\in\{3,5\}\) (20 scenarios per cell). All known-order random-multipath scenarios enforce a minimum inter-path delay spacing of approximately \(1.2W_\tau\). To isolate the roles of candidate validation (V), resolution pruning (P), and residual redetection (R), a full \(2^3\) factorial ablation tests all eight on/off module combinations across 360 scenarios. A separate 2300-run validation study then compares the pruned--redetected (PR) simplification against the full VPR pipeline across one-dimensional near--far, two-dimensional known- and unknown-order, and Watermark communication benchmarks while holding all thresholds and iteration limits fixed.

Throughout the synthetic benchmarks, the independent statistical unit is a base scenario (defined by its true path parameters, noise seed, and probe). We compute performance differences between methods within each scenario before summarizing paired statistics across scenarios; multiple paths within a single scenario are never treated as independent observations. Similarly, because the near--far sweeps in Figs.~\ref{fig:7} and~\ref{fig:10}(a) scale the weak-path amplitude across NFR levels while reusing the same base noise realization, different NFR points within a base scenario represent paired conditions rather than independent samples.

In the measured Watermark communication replay, we pair estimators within each combination of channel, snapshot, and SNR. Testing snapshots 0--29 of channels NOF1\_001 and KAU1\_001 across SNRs of 10, 20, and 30 dB produces 180 paired frames organized into 60 snapshot groups. To compute overall summary statistics without pseudoreplication across SNRs, we first average each metric (or paired difference) across the three SNRs within each snapshot, summarize across the 30 snapshots of each channel, and finally weight the two channels equally. Each point in Fig.~\ref{fig:13} averages over the 30 snapshots of the corresponding channel at that SNR. Statistical inferences from this replay apply strictly to the recorded propagation conditions of NOF1\_001 and KAU1\_001. For the 50 measured records from the UWA Channel Library, snapshots with matching indices share a standardized noise sequence; we therefore report those 50 fitting residuals descriptively rather than as independent random samples.

\subsection{Scenarios and Experimental Overview}\label{sec:6.2}

Table~\ref{tab:2} summarizes the ten experimental investigations, their sampling units, total execution counts, and statistical inference scopes.

\begin{table*}[!t]
\caption{Summary of experimental investigations}\label{tab:2}
\centering\footnotesize
\setlength{\tabcolsep}{3pt}
\begin{tabularx}{\textwidth}{YYYYY}
\toprule

Investigation & Configuration & Sample unit & Total runs & Inference scope \\

\midrule

Factorial step ablations & Known/unknown order, high-order multipath, noise only; 8 V/P/R variants & 360 scenarios & 2880 & Paired differences within parameter cells \\

1-D unknown order & \(L\in\{0,1,2,3,5,8\}\), SNR \(\in\{0,10,20\}\) dB & 1080 scenarios & 4320 & Overall, stratified cells, difficult subset \\

2-D NOMP comparison (known order) & Random multipath and near--far two-path & 280 scenarios & 840 & Overall performance and scenario subsets \\

2-D SNR sweep (known order) & \(L\in\{3,5\}\), SNR \(\in\{0,10,20,30\}\) dB & 160 scenarios & 320 & Performance trends across SNRs \\

2-D unknown order & \(L\in\{0,1,2,3,5,8\}\), SNR \(\in\{0,10,20\}\) dB & 1080 scenarios & 2160 & Overall, stratified cells, difficult subset \\

Optional near--far replacement & 8 near--far cases; random \(L\in\{5,8\}\), SNR \(\in\{0,10,20\}\) dB & 760 scenarios & 2280 & Near--far recovery and stability checks \\

PR simplification ablation & 1-D extreme near--far, 2-D scenarios, Watermark BER & 2300 runs & 4960 & Validation impact on boundary stability \\

Measured Watermark BER & 2 Watermark channels, 3 SNRs, 30 snapshots & 180 paired frames & 1800 & Stratified snapshot analysis on recorded data \\

Repeated runtime benchmarks & 25 ms LFM, 10 dB SNR, \(L\in\{1,2,3,5,8\}\) & 50 paired runs/method & 250 & Single-thread execution time and memory \\

PR runtime comparison & 1-D \(L=5,8\), 2-D \(L=5\) & 60 repeated runs & 60 & Isolated single-thread execution \\

\bottomrule

\end{tabularx}

\end{table*}

\subsection{Baselines and Computational-Cost Measurement}\label{sec:6.3}

We benchmark RC-SAGE--Brent against four established estimators: OMP \cite{TroppGilbert2007OMP}, WRELAX \cite{LiWu1998WRELAX}, off-grid SBL \cite{Tipping2001SBL,YangXieZhang2013OGSBL}, and two-dimensional NOMP \cite{Mamandipoor2016NOMP}. We also include fixed coarse-grid (Coarse) and fine-grid (Fine) matched filters as reference baselines, along with standalone SAGE--Brent refinement as an ablation control. In all one-dimensional unknown-order benchmarks, every estimator uses the exact same median-based noise variance estimator~\eqref{eq:19}, effective cell count \(M_{\mathrm{eff}}\), minimum delay separation \(0.8W_\tau\), and maximum order \(L_{\max}=10\). While RC-SAGE--Brent generates an initial candidate list and applies CFAR thresholding across its validation, pruning, and redetection stages, the sequential baselines (OMP, WRELAX, and SBL) apply the same CFAR threshold as their forward stopping rule, terminating as soon as the next candidate falls below \(\gamma_r\). Note that the curve labeled SBL-MP in Fig.~\ref{fig:6} represents our single-snapshot delay off-grid SBL baseline rather than the multisnapshot MMV algorithm of Yang and Choo \cite{YangChoo2025TimeDomainOGSBL}.

Specifically, our OMP baseline performs joint least-squares gain refitting after each path admission, allows up to 60 objective evaluations for single-path Gauss--Newton refinement, and executes one final refinement pass. WRELAX initializes each new path via a global one-dimensional grid search, cyclically relaxes all active path delays using Brent's method for up to eight cycles (or until delay changes fall below \(10^{-3}/f_s\)), and finishes with a joint least-squares gain fit. SBL operates on a dictionary of 64 delay candidates spaced at \(T_s/4\) for up to 200 EM iterations, stopping early after 10 iterations if relative hyperparameter changes drop below \(10^{-2}\), and then applies four Brent refinement passes to the surviving support. Supplementary Table S3 lists the full baseline configurations.

For two-dimensional delay--scale benchmarks, the NOMP baseline searches the identical coarse grid, oversamples each parameter axis by a factor of four around the detected peak, and applies two-dimensional Newton refinements, cyclic Newton relaxation, and joint least-squares gain updates \cite{Mamandipoor2016NOMP}. It uses neither Brent search nor closed-loop support pruning and redetection. In unknown-order trials, two-dimensional NOMP shares the exact same \(M_{\mathrm{eff}}\), \(P_{\mathrm{FA}}=10^{-3}\), \(L_{\max}=10\), and normalized minimum separation \(d_{\min}=0.6\) as RC-SAGE--Brent. Each algorithm retains its native gain update: pathwise conditional least squares for RC-SAGE--Brent versus full-support joint least squares for NOMP. Because a fair two-dimensional off-grid SBL baseline would require bespoke two-dimensional dictionary pruning and hyperprior tuning, we leave two-dimensional SBL comparisons to future work.

We also compare RC-SAGE--Brent against the recently proposed fast variational Bayes (FVB) and sparse variational Bayes (SVB) estimators developed by Halder and Murthy for doubly spread channels \cite{HalderMurthy2025DSSpread}. Retaining their public MATLAB inference core, we constructed a sample-domain adaptation whose dictionary columns match the exact continuous time-scaled waveforms of~\eqref{eq:1}. We evaluate this adaptation on the 200 frozen 25 ms observations of Table~\ref{tab:7}. Both FVB and SVB use a \(50\times5\) delay--scale dictionary, initialize six paths via OMP, and update the 13 strongest dictionary columns per iteration. To ensure an equal-capacity, equal-budget comparison, RC-SAGE--Brent, FVB, and SVB all return at most six paths within a hard 15 s wall-clock deadline and apply identical backward deletion, joint least-squares refitting, and model-order selection (fixed penalty for PN, residual-adaptive penalty for LFM). We report reconstruction accuracy for both the full 250-column VB posterior and the pruned six-path sparse output. While this protocol equalizes the physical signal model, output sparsity, and runtime budget, we emphasize that it evaluates a sample-domain adaptation of FVB and SVB rather than their native multicarrier OFDM formulation.

For unknown-order benchmarks, we record the end-to-end execution time encompassing coarse discovery, continuous refinement, support updates, and order selection across all 1080 scenarios as well as on the difficult subset (\(L\ge3\), \(\mathrm{SNR}\ge10\ \mathrm{dB}\)). For dedicated known-order complexity profiling (Section~\ref{sec:7.5}), we execute each run in an isolated process with BLAS, MKL, and OpenMP pinned to a single CPU thread. After discarding one warm-up call, we record the median runtime, objective evaluation count, and peak RSS increment across 10 repeated calls per method and path count. The optional replacement and PR simplification timing tests follow this identical single-thread protocol. (The 280-scenario two-dimensional known-order accuracy benchmark used multi-threaded worker pools to evaluate estimation accuracy only and is excluded from runtime comparisons.)

To summarize paired runtimes between method A and method B across scenarios \(i\), we compute the per-scenario runtime ratio \(r_i=t_{A,i}/t_{B,i}\) and report the median paired ratio \(R_t=\operatorname{median}_i r_i\), along with the 10th and 90th empirical percentiles (P10 and P90). A median ratio \(R_t<1\) indicates that method A runs faster than method B on a paired per-scenario basis. Note that the median of ratios \(R_t\) generally differs from the ratio of separate median runtimes; we never pair timing records collected under different threading rules or algorithmic stages.

Finally, to test whether two-dimensional unknown-order estimation scales to shorter probes and harsher propagation conditions under a strict real-time budget, we conducted a 200-scenario production validation using 25 ms probes (100 PN and 100 LFM scenarios). Each waveform suite spans SNRs of 10, 15, 20, and 25 dB (25 scenarios per SNR), covers true orders \(L\in\{1,\dots,5\}\) uniformly, and stratifies scenarios equally across five regimes: ordinary random multipath, extreme near--far, closely spaced dense multipath, parameter-boundary arrivals, and mixed stress. Operating blind to the true order with a cap of six output paths, the estimator must complete dictionary setup, closed-loop estimation, and final order selection within a hard 15 s deadline; if a run reaches 15 s, we output the latest completed iterate. As noted in Section~\ref{sec:4.3}, we calibrated the adaptive LFM penalty~\eqref{eq:22c} on a separate 100-scenario development set before running the 100-scenario validation set with disjoint random seeds, and we include the fixed penalty \(\lambda=1\) on the same LFM validation candidates as an ablation control. Prior to validation, we set engineering target thresholds of 95\% order accuracy for PN, 85\% order accuracy for LFM, and at most 5\% severe error rate; after observing 84\% LFM order accuracy on the validation set, we relaxed the LFM acceptance criterion to 80\%, which we explicitly report as a post hoc engineering adjustment.

\section{Experimental Results}\label{sec:7}

\subsection{Continuous Refinement and Known-Order Multipath Estimation}\label{sec:7.1}

\begin{figure*}[!t]
\centering
\includegraphics[width=\textwidth,height=.72\textheight,keepaspectratio]{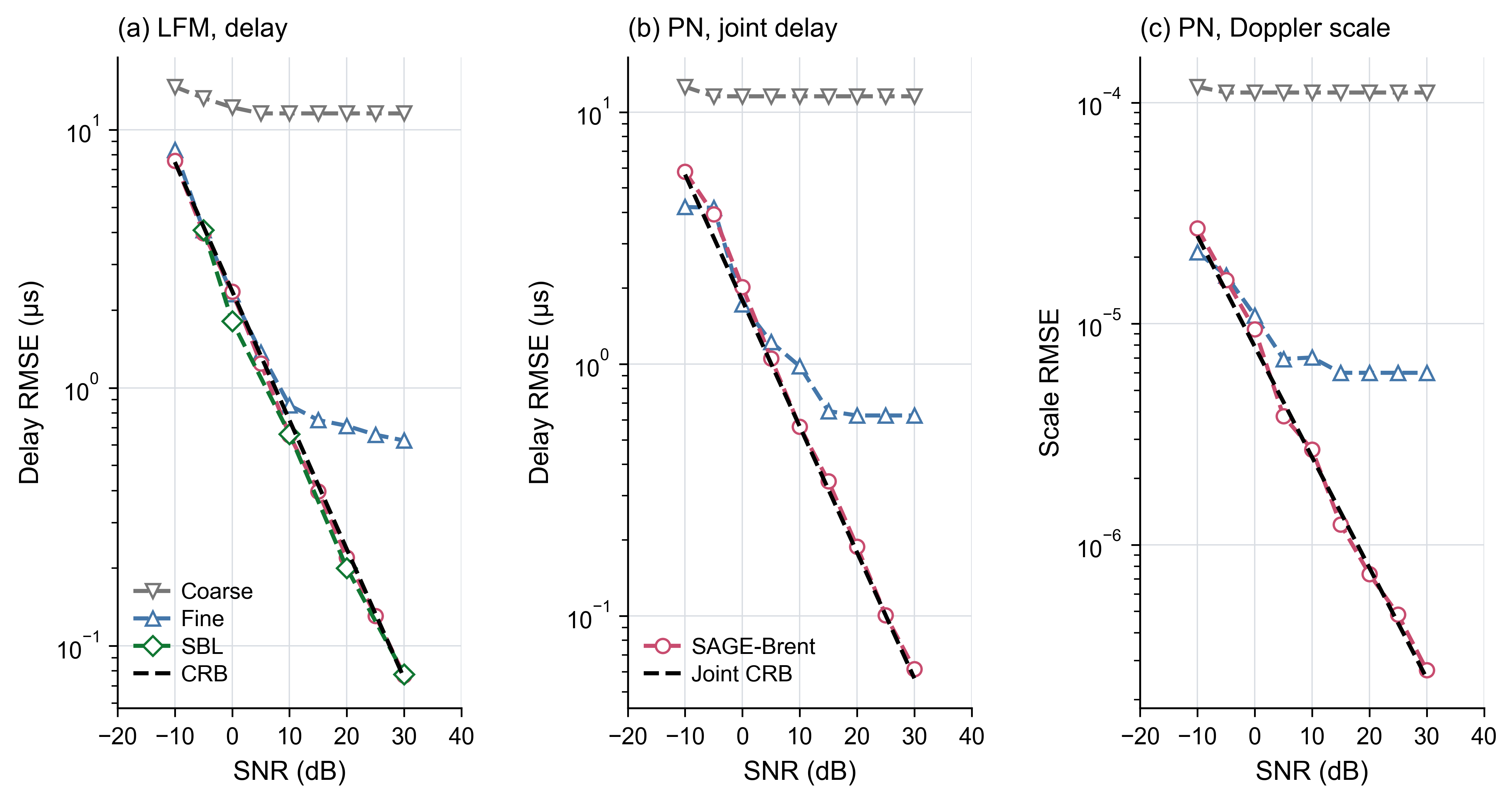}
\caption{Single-path parameter accuracy: (a) LFM delay RMSE; (b) delay RMSE for joint PN estimation; (c) PN scale RMSE.}\label{fig:4}

\end{figure*}

Fig.~\ref{fig:4}(a) compares one-dimensional delay estimation across discrete grids, SBL, and the proposed continuous refinement. Figs.~\ref{fig:4}(b) and (c) benchmark two-dimensional grids, alternating SAGE--Brent refinement, and the joint Cram\'er--Rao bound (CRB) for a single PN path. Both two-dimensional evaluations share identical grid configurations and refinement settings, without two-dimensional SBL or OGSBL. At 20 dB SNR, one-dimensional continuous refinement achieves a delay RMSE of 0.220 \(\mu\mathrm{s}\), whereas the fine grid stalls at 0.713 \(\mu\mathrm{s}\). In two dimensions, the proposed refinement reaches a delay RMSE of 0.189 \(\mu\mathrm{s}\), compared with 0.625 \(\mu\mathrm{s}\) for the fine grid. Across the tested SNR range, continuous local optimization removes the quantization error floor imposed by discrete grid resolution.

\begin{figure*}[!t]
\centering
\includegraphics[width=\textwidth,height=.72\textheight,keepaspectratio]{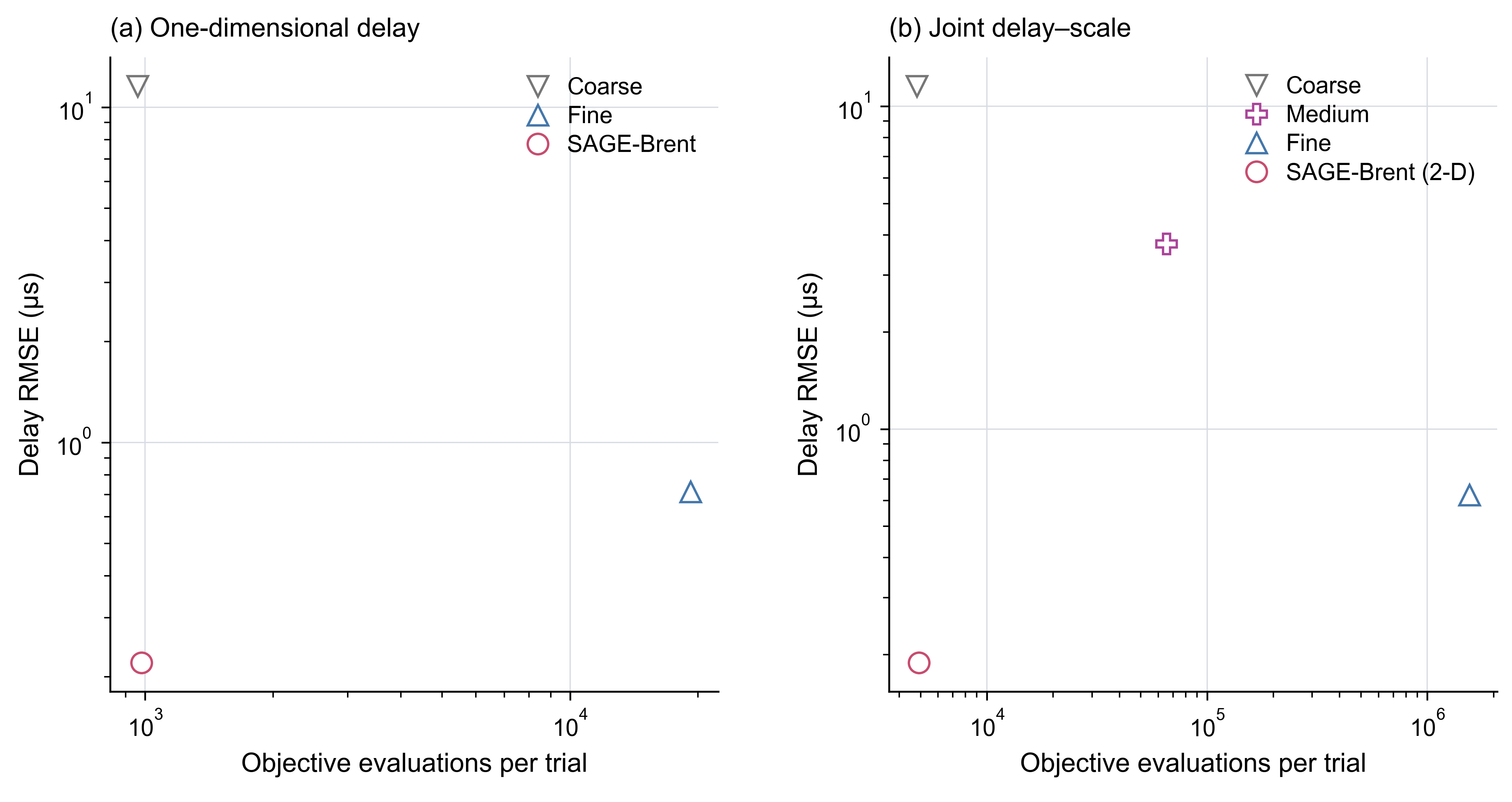}
\caption{Accuracy versus objective evaluations at 20 dB: (a) 1-D delay refinement; (b) 2-D delay--scale refinement.}\label{fig:5}

\end{figure*}

Fig.~\ref{fig:5} plots parameter RMSE against the number of objective evaluations at 20 dB SNR, including the coarse grid as a reference baseline. In Fig.~\ref{fig:5}(b), SAGE--Brent denotes Algorithm~\ref{alg:1}, which updates delay at a fixed scale and then Doppler scale at the updated delay. In one dimension, continuous refinement converges after approximately 981 objective evaluations, whereas the fine grid requires 19201 evaluations. In two dimensions, alternating refinement requires roughly 4909 evaluations, whereas an exhaustive two-dimensional fine grid requires \(1.56\times10^6\) evaluations. The computational cost of a two-dimensional grid scales with the product of the grid sizes along both axes. By updating the two coordinates sequentially, alternating search reduces objective evaluations by more than two orders of magnitude while achieving superior parameter accuracy. Section~\ref{sec:7.5} reports measured wall-clock execution times.

\begin{figure*}[!t]
\centering
\includegraphics[width=\textwidth,height=.72\textheight,keepaspectratio]{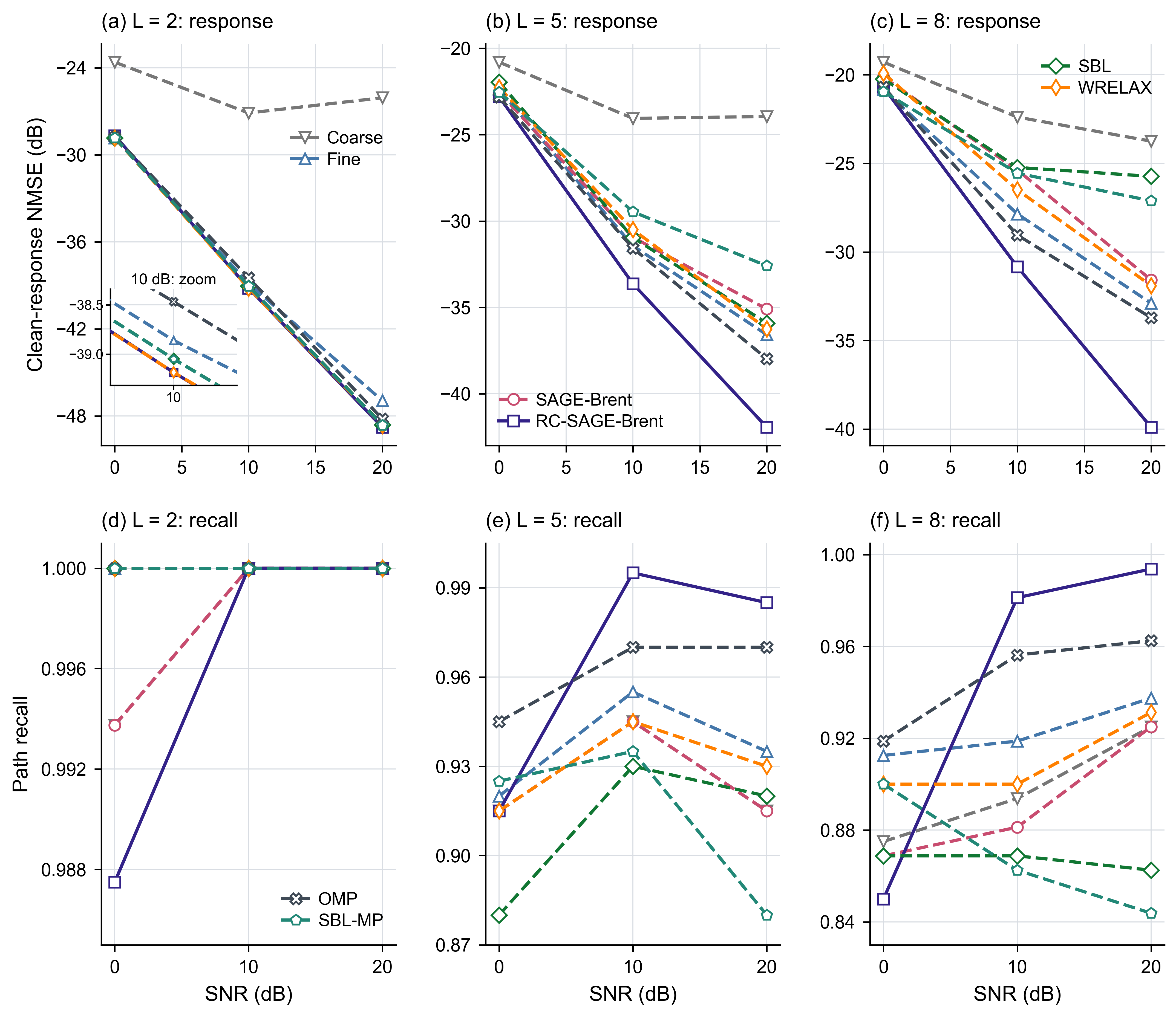}
\caption{Known-order random multipath: (a)--(c) response NMSE for \(L=2,5,8\); (d)--(f) corresponding path recall. The upper-right inset in (a) enlarges the region at SNR=10 dB.}\label{fig:6}

\end{figure*}

Fig.~\ref{fig:6} evaluates paired random-multipath scenarios with a minimum inter-path delay separation of approximately \(1.2W_\tau\). The recall curves in the lower row record parameter-matched arrivals without an amplitude threshold, and the vertical axes highlight differences in the high-recall regime. The inset in Fig.~\ref{fig:6}(a) magnifies the curves around 10 dB SNR without adding new simulation points. At 20 dB SNR with \(L=5\), RC-SAGE--Brent achieves a response NMSE of \(-41.94\) dB and a recall of 0.985. Under the same conditions, standalone refinement achieves \(-35.09\) dB and 0.915, whereas OMP yields \(-37.97\) dB and 0.970. For \(L=8\), RC-SAGE--Brent reaches \(-39.90\) dB NMSE and 0.994 recall, outperforming standalone refinement (\(-31.58\) dB, 0.925) and OMP (\(-33.71\) dB, 0.960). Performance gains concentrate in moderate- and high-order channels, whereas all continuous methods approach estimation saturation in low-order settings.

\begin{figure*}[!t]
\centering
\includegraphics[width=\textwidth,height=.72\textheight,keepaspectratio]{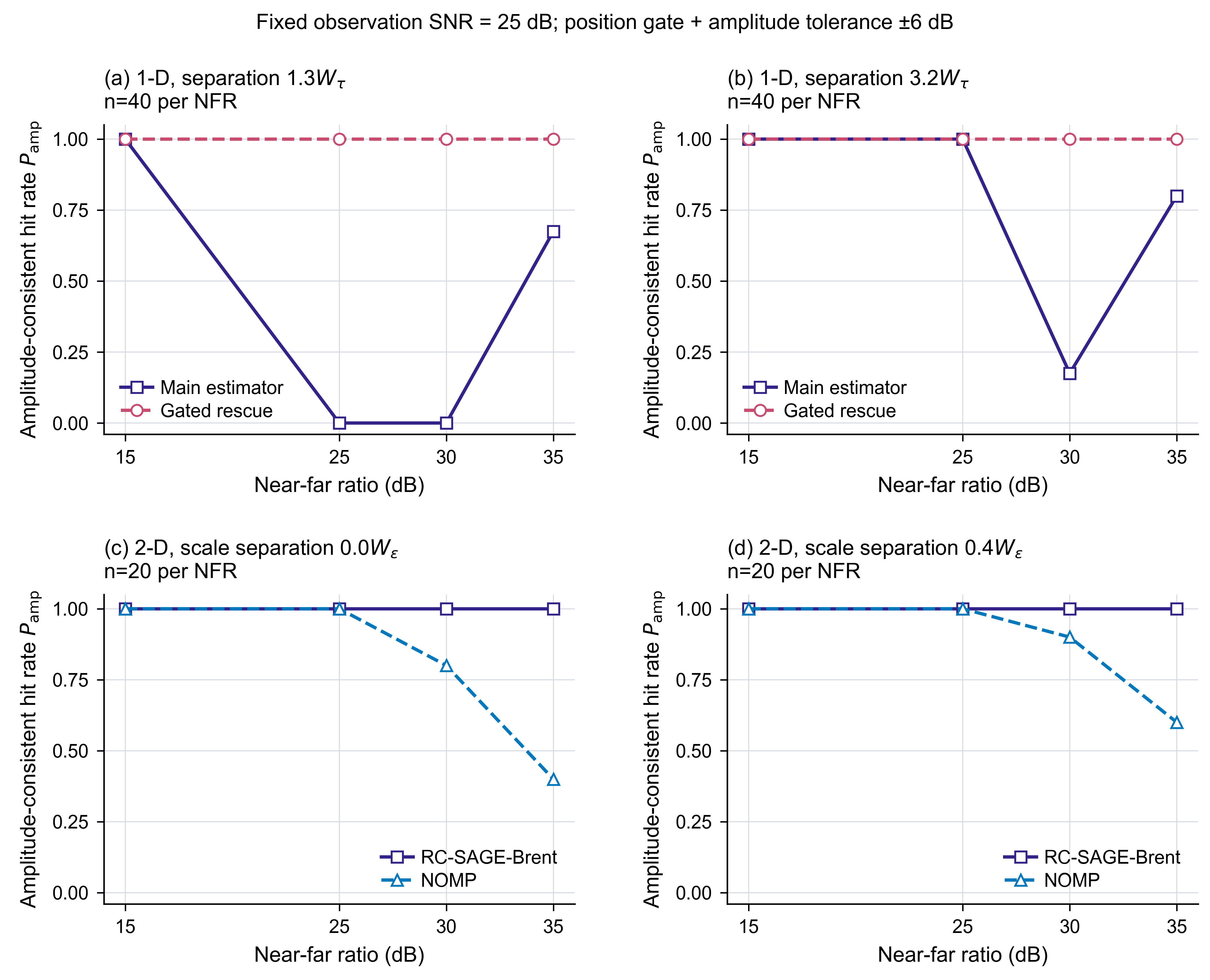}
\caption{Amplitude-consistent near--far two-path detection: (a)(b) 1-D delay spacings \(1.3W_\tau\), \(3.2W_\tau\); (c)(d) 2-D scale spacings 0, \(0.4W_\epsilon\).}\label{fig:7}

\end{figure*}

Fig.~\ref{fig:7} sweeps the near--far ratio (NFR) from 15 to 35 dB at an overall SNR of 25 dB. At an NFR of 35 dB, the weak path carries only \(1/56.2\) of the strong-path amplitude. We evaluate amplitude-consistent detection via~\eqref{eq:10a}, requiring the estimated path to match both location tolerances and lie within \(\pm6\) dB of the true weak-path gain. Each one-dimensional delay spacing uses 40 base scenarios, and each two-dimensional scale spacing uses 20. Within each base scenario, all four NFR levels share identical noise realizations, representing paired test conditions. At a delay spacing of \(1.3W_\tau\), standard one-dimensional RC-SAGE--Brent detects the weak arrival at rates of 1.000, 0, 0, and 0.675 for NFRs of 15, 25, 30, and 35 dB, respectively. At a wider spacing of \(3.2W_\tau\), the detection rates are 1.000, 1.000, 0.175, and 0.800. Among correctly matched detections, the median amplitude estimation errors remain within 0.082 dB and 0.029 dB for the two spacings. In control trials without a weak path, the estimator produced zero false alarms inside the matching gate.

Stagewise diagnostics reveal why weak-path detection rebounds at 35 dB NFR. At a delay spacing of \(1.3W_\tau\) and NFRs of 25, 30, and 35 dB, coarse detection identifies the strong path and one spurious residual peak, missing the true weak arrival. After continuous refinement, the median ratios of spurious-candidate energy to the validation threshold are 5.286, 2.277, and 0.815 across the three NFRs. At 25 and 30 dB NFR, the spurious candidate exceeds the threshold and consumes the final slot of the two-path budget. The residual search subsequent to this step detects the weak arrival, but the estimator cannot admit it because the path budget is full. At 35 dB NFR, the spurious candidate falls below the validation threshold in 27 out of 40 trials. Candidate validation rejects the spurious peak, which creates a vacancy for initial residual redetection to admit the true weak arrival, producing an empirical detection rate of 0.675 (27/40). At \(3.2W_\tau\) spacing, the energy ratios at 30 and 35 dB NFR are 1.730 and 0.711, and validation rejects the spurious candidate in 7 and 32 of 40 trials, matching the observed detection rates of 0.175 and 0.800. Across all 320 diagnostic runs, path decisions before and after cyclic refinement were identical; therefore, the rebound originates in the validation and initial-redetection stages.

This rebound does not imply that weak paths become easier to detect at higher dynamic ranges. Rather, increasing the NFR scales the weak path, the spurious peak, and the cancellation residual simultaneously. The weak path enters the model state only when the spurious peak falls below the detection threshold, freeing a slot in the candidate budget. We confirmed this mechanism with two noiseless control experiments. In those trials, the spurious peak persisted at 35 dB NFR; however, replacing coarse candidate parameters with true values restored weak-path detection. Thus, coarse candidate discovery and the path budget jointly cause the intermediate failure.

The optional known-order replacement mechanism addresses this capacity-constrained edge case. The routine executes at most once per trial, activating only when three criteria hold simultaneously: the active path count equals the model order, the estimated gain range exceeds 25 dB, and the residual peak exceeds the detection threshold. The algorithm accepts the replacement only if it reduces the residual sum of squares by at least 3\%. Enabling replacement restores weak-path detection to 1.000 across all conditions in Figs.~\ref{fig:7}(a) and (b), while control runs without weak paths remain at zero. Standard RC-SAGE--Brent disables this replacement step. In the two-dimensional trials of Figs.~\ref{fig:7}(c) and (d), standard RC-SAGE--Brent achieves 1.000 detection across all four NFRs without replacement, whereas NOMP degrades steadily as the NFR increases. Therefore, the primary two-dimensional results do not depend on the replacement routine.

\subsection{Path Selection, Updating, and Unknown-Order Comparisons}\label{sec:7.2}

\begin{figure*}[!t]
\centering
\includegraphics[width=\textwidth,height=.72\textheight,keepaspectratio]{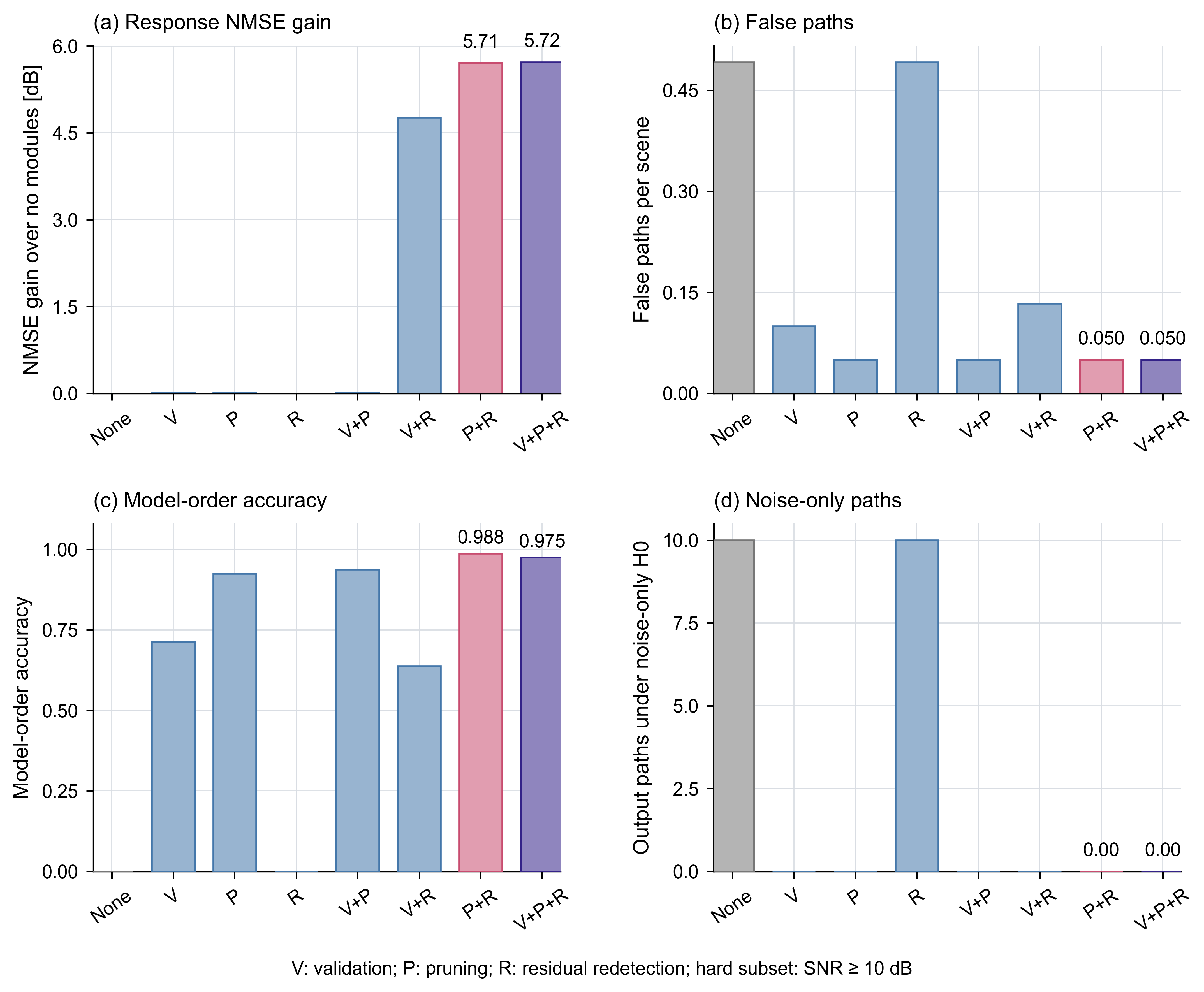}
\caption{Controlled comparisons of all V/P/R combinations: (a) response NMSE gain; (b) known-order spurious-path count; (c) unknown-order accuracy; (d) estimated path count under noise only.}\label{fig:8}

\vspace{-2pt}
\end{figure*}

Fig.~\ref{fig:8} evaluates all eight combinations of candidate validation (V), resolution pruning (P), and residual redetection (R). The known-order panels and nonempty unknown-order panels use the difficult subset of 240 scenarios with \(L\in\{5,8\}\) and SNR \(\ge10\) dB. Fig.~\ref{fig:8}(d) reports the 240 noise-only scenarios. Residual redetection requires validation and pruning to clear false peaks first. Without them, redetection alone performs identically to the baseline with all three steps disabled. When we add redetection to validation and pruning (VPR versus VP), response NMSE improves by 5.71 dB, and path recall rises from 0.921 to 0.992. Resolution pruning provides the largest improvement in order accuracy. Adding pruning to validation and redetection (VPR versus VR) increases unknown-order accuracy by 0.338 and path \(F_1\) by 0.034, while reducing the severe failure rate by 0.100. In known-order trials, adding pruning improves NMSE by 0.95 dB and removes 0.083 spurious paths per scenario.

Candidate validation alone suppresses known-order false paths from 0.492 to 0.100 per scenario and raises unknown-order accuracy from 0 to 0.713. Once pruning and redetection are active, however, adding validation changes unknown-order accuracy by only \ensuremath{-}0.013. In these high-order random channels, validation and iterative pruning serve overlapping functions. The PR and VPR configurations exhibit nearly identical performance in Fig.~\ref{fig:8}. For PR and VPR, known-order NMSE values are \ensuremath{-}37.160 dB and \ensuremath{-}37.172 dB, path recall is 0.9919 for both, spurious paths average 0.050 per scenario for both, and unknown-order accuracy reaches 0.9875 and 0.9750, respectively. Pruning and redetection therefore deliver most of the performance gain in ordinary random multipath. Validation proves essential when preserving weak paths under extreme near--far conditions.

To evaluate whether validation can be omitted safely, we conducted paired trials comparing PR with the full VPR architecture. PR omits initial candidate validation but retains resolution pruning and residual redetection. Table~\ref{tab:3} reports the paired performance differences, where positive values favor PR.

\begin{table}[!t]
\caption{Paired performance differences for PR candidate simplification}\label{tab:3}
\centering\footnotesize
\setlength{\tabcolsep}{3pt}
\begin{tabularx}{\columnwidth}{YY}
\toprule

Scenario and metric & PR improvement \\

\midrule

1-D random unknown order: order accuracy & +0.0056 \\

1-D extreme near--far: weak-path detection & \ensuremath{-}0.0969 \\

1-D extreme near--far: response NMSE & \ensuremath{-}1.7689 dB \\

2-D known order: path recall & \ensuremath{-}0.0036 \\

2-D near--far: weak-path detection & \ensuremath{-}0.0167 \\

2-D unknown order: order accuracy & \ensuremath{-}0.0009 \\

Watermark CFAR10: BER & \ensuremath{-}0.00052 \\

Watermark CFAR20-JLS: BER & +0.00014 \\

\bottomrule

\end{tabularx}

\end{table}

Candidate validation provides critical protection under extreme near--far dynamics and parameter boundaries. By contrast, performance differences in standard random multipath, two-dimensional unknown-order estimation, and overall communication BER remain modest. The most pronounced one-dimensional gap appears at 35 dB NFR with delay spacing \(1.3W_\tau\), where weak-path detection rates for VPR and PR are 0.637 and 0.037, respectively. In two-dimensional known-order estimation at 25 dB NFR with scale separation \(0.4W_\epsilon\), VPR achieves a detection rate of 0.800 compared with 0.700 for PR, and omitting validation degrades response NMSE by 3.113 dB. This degradation occurs because unvalidated spurious candidates consume the limited path budget early in the search. Although subsequent pruning can remove false paths, recovering masked weak arrivals depends on remaining iterations and redetection opportunities.

In summary, pruning removes false paths to stabilize model order, creating vacancies for residual redetection to recover genuine weak arrivals. While validation and pruning overlap functionally in standard channels, validation protects weak arrivals under severe near--far dynamics. RC-SAGE--Brent therefore retains the complete VPR pipeline to balance nominal accuracy with boundary robustness. This architecture differs fundamentally from optional replacement: validation applies a threshold check before candidate admission, whereas replacement tests an exchange only after the candidate budget is exhausted.

\begin{figure*}[!t]
\centering
\includegraphics[width=\textwidth,height=.72\textheight,keepaspectratio]{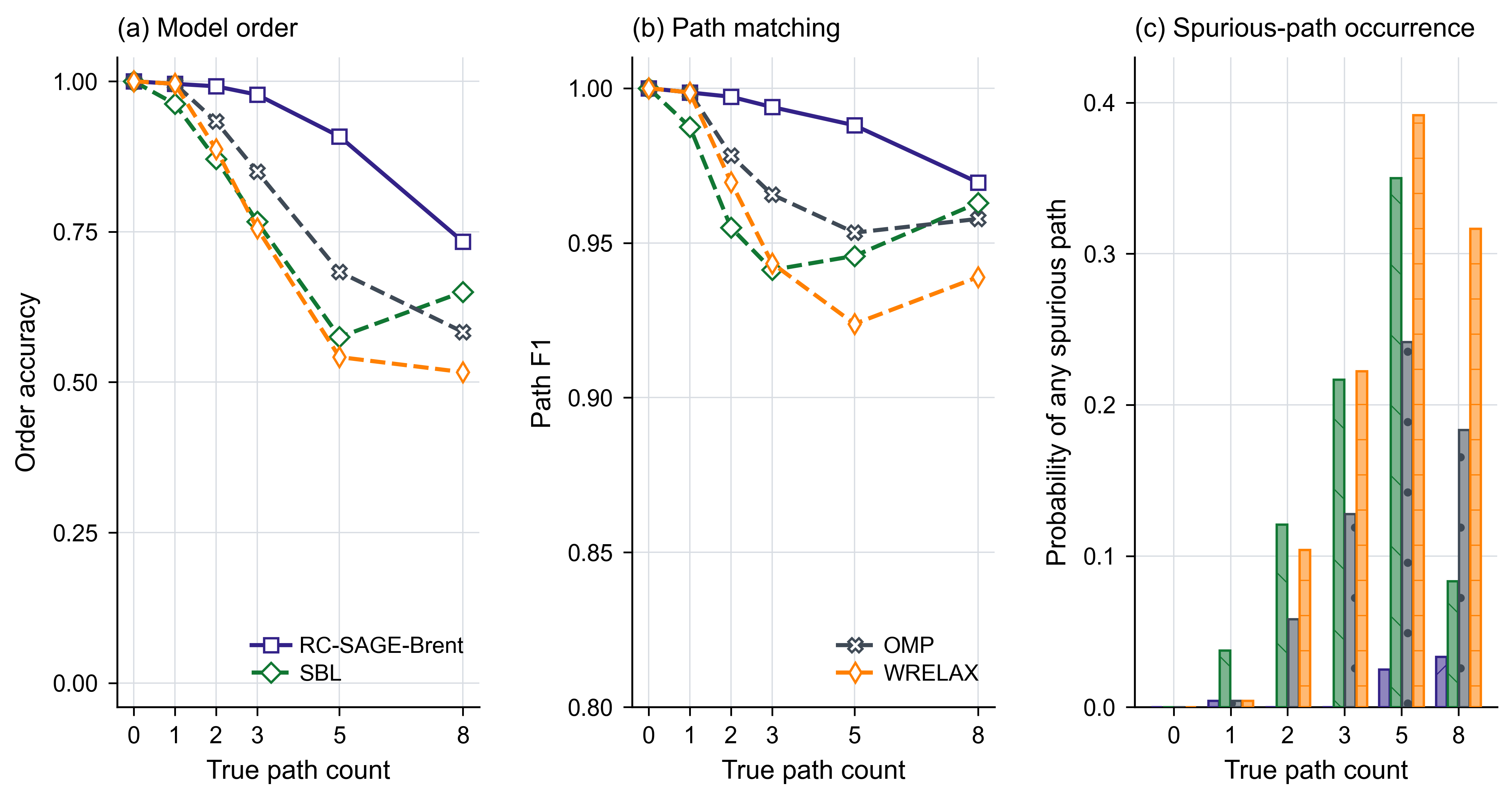}
\caption{1-D unknown-order estimation: (a) order accuracy; (b) path F1; (c) scenario-level spurious-path occurrence rate.}\label{fig:9}

\end{figure*}

Fig.~\ref{fig:9} summarizes one-dimensional performance across true path counts \(L\in\{0,1,2,3,5,8\}\), weighting the three SNR levels equally. Tables~\ref{tab:4} and~\ref{tab:5} report overall and difficult-subset statistics, respectively. In Fig.~\ref{fig:9}(c), the spurious-path rate \(P_{\mathrm{spur}}=N^{-1}\sum_i\mathbb 1\{N_{\mathrm{FP},i}>0\}\) measures the proportion of trials containing false paths, separating overestimation from missed detections. At \(L=5\), spurious-path rates for RC-SAGE--Brent, OMP, SBL, and WRELAX are 2.50\%, 24.17\%, 35.00\%, and 39.17\%, respectively. These rates correspond to mean spurious counts of 0.042, 0.458, 0.533, and 0.833 paths per scenario.

\begin{table*}[!t]
\caption{Overall unknown-order performance}\label{tab:4}
\centering\footnotesize
\setlength{\tabcolsep}{3pt}
\begin{tabularx}{\textwidth}{YYYYYY}
\toprule

Method & Order accuracy & Order MAE & F1 & Severe failure rate & Median full estimation time/ms \\

\midrule

RC-SAGE--Brent & 0.969 & 0.043 & 0.995 & 0.007 & 1078.8 \\

SBL & 0.857 & 0.248 & 0.969 & 0.058 & 1850.0 \\

OMP & 0.901 & 0.166 & 0.982 & 0.039 & 628.8 \\

WRELAX & 0.856 & 0.235 & 0.972 & 0.060 & 457.6 \\

\bottomrule

\end{tabularx}

\end{table*}

\begin{table*}[!t]
\caption{Scenario-wise paired differences for RC-SAGE--Brent}\label{tab:5}
\centering\footnotesize
\setlength{\tabcolsep}{3pt}
\begin{tabularx}{\textwidth}{YYYYYY}
\toprule

Scope & Baseline & Order accuracy gain & NMSE gain/dB & Severe failure reduction & Time ratio \\

\midrule

Overall & SBL & 0.111 & 3.947 & 0.051 & 0.623 \\

Overall & OMP & 0.068 & 2.471 & 0.031 & 1.831 \\

Overall & WRELAX & 0.113 & 3.237 & 0.053 & 2.521 \\

Difficult subset & SBL & 0.329 & 7.493 & 0.142 & 0.450 \\

Difficult subset & OMP & 0.237 & 6.267 & 0.108 & 1.008 \\

Difficult subset & WRELAX & 0.396 & 8.119 & 0.217 & 0.487 \\

\bottomrule

\end{tabularx}

\end{table*}

On the difficult subset (240 scenarios with \(L\ge3\) and SNR \(\ge10\) dB), RC-SAGE--Brent achieves an order accuracy gain of 0.237 over OMP and improves response NMSE by 6.267 dB, while maintaining an identical median runtime (paired time ratio 1.008). These results confirm that the closed-loop architecture improves order selection without adding computational overhead. At \(L=8\) and 0 dB SNR, all algorithms become detectability-limited, and RC-SAGE--Brent achieves an order accuracy of 0.25. Across all 240 noise-only trials, none of the four methods generated false alarms.

\subsection{Two-Dimensional Delay--Scale Estimation}\label{sec:7.3}

\begin{figure*}[!t]
\centering
\includegraphics[width=\textwidth,height=.72\textheight,keepaspectratio]{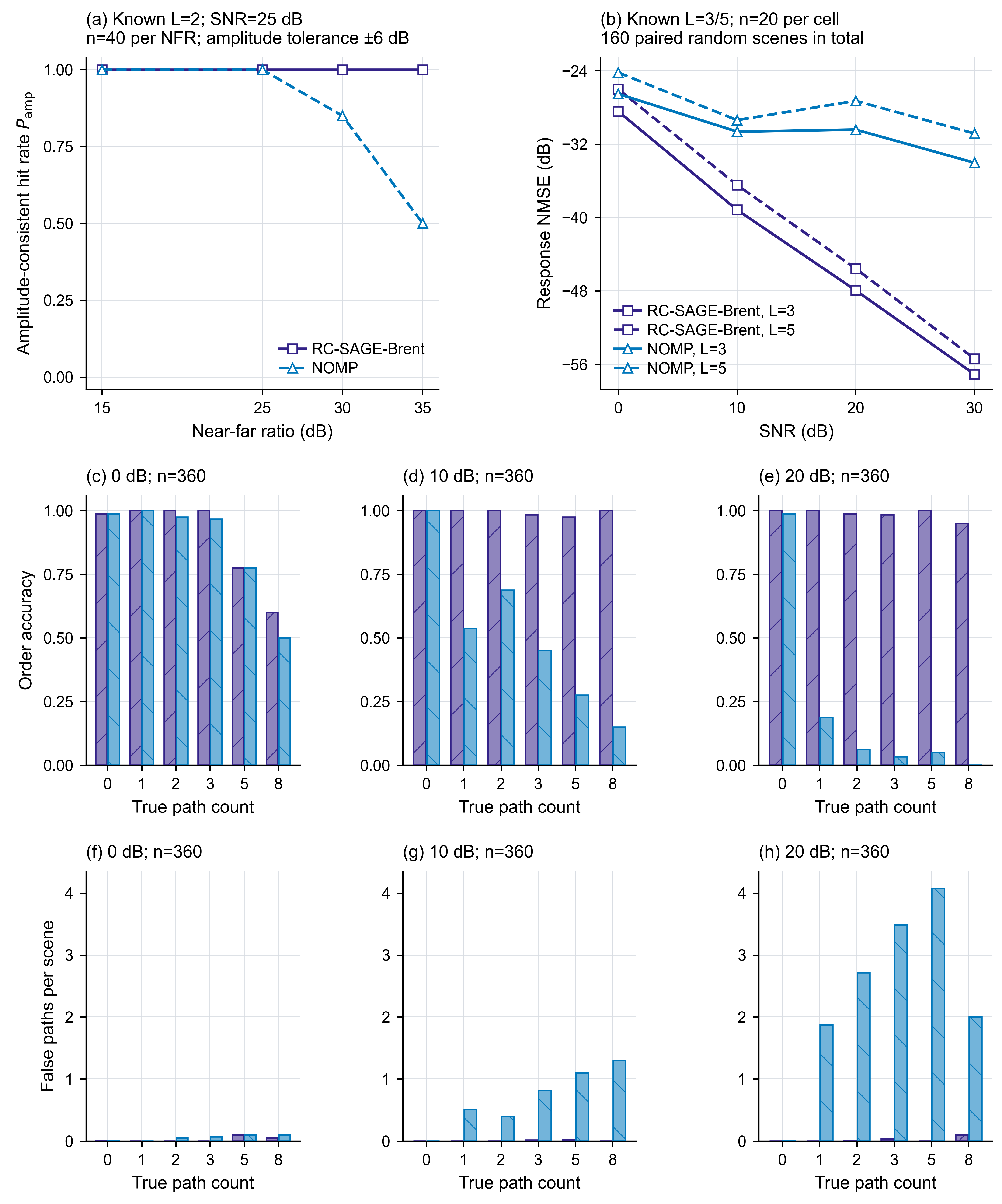}
\caption{2-D estimation: (a) near--far two-path detection; (b) known-order response NMSE; (c)--(e) order accuracy at 0, 10, 20 dB; (f)--(h) corresponding spurious paths per scenario.}\label{fig:10}

\end{figure*}

Fig.~\ref{fig:10} summarizes three distinct two-dimensional experiments. Panel (a) presents a PN near--far diagnostic with two known paths at 25 dB overall SNR. Panel (b) illustrates a PN known-order SNR sweep for \(L\in\{3,5\}\) with 20 paired scenarios per SNR and path count. Panels (c)--(h) report unknown-order results across the 1080 scenarios. Supplementary Table S2 documents the sample counts for each panel. In Fig.~\ref{fig:10}(b), RC-SAGE--Brent improves response NMSE over NOMP by 1.90 dB for \(L=3\) and by 1.82 dB for \(L=5\) at 0 dB SNR. At 30 dB SNR, these NMSE gains increase to 23.05 dB and 24.54 dB, respectively. The NOMP error floor stalls at moderate and high SNRs, whereas the error of RC-SAGE--Brent continues to decline across the tested range. Because both methods include continuous coordinate refinement, these results benchmark complete estimators under the wideband delay--scale model rather than isolated optimizers.

Across the 280 paired known-order scenarios, RC-SAGE--Brent achieves a recall of 0.986, an \(F_1\) score of 0.986, and a response NMSE of \(-47.29\) dB. NOMP achieves 0.967 recall, 0.967 \(F_1\), and \(-34.16\) dB NMSE, while the coarse grid achieves 0.846 recall, 0.846 \(F_1\), and \(-21.58\) dB NMSE. RC-SAGE--Brent exceeds NOMP by 0.018 in recall and by 13.12 dB in response NMSE. In the random-multipath subset, the recall advantage narrows to 0.001, indicating that both algorithms detect paths at similar rates in benign channels; however, RC-SAGE--Brent maintains a 12.12 dB lead in response NMSE.

\begin{table*}[!t]
\caption{2-D unknown-order performance overall and on the difficult subset}\label{tab:6}
\centering\footnotesize
\setlength{\tabcolsep}{3pt}
\begin{tabularx}{\textwidth}{YYYYYYY}
\toprule

Scope & Method & Order accuracy & F1 & Spurious paths/scenario & Severe failure rate & Median time/ms \\

\midrule

Overall & RC-SAGE--Brent & 0.979 & 0.997 & 0.012 & 0.005 & 6790.8 \\

Overall & NOMP-2D-CFAR & 0.609 & 0.877 & 0.914 & 0.253 & 1659.2 \\

Difficult subset & RC-SAGE--Brent & 0.983 & 0.998 & 0.025 & 0.008 & 8171.8 \\

Difficult subset & NOMP-2D-CFAR & 0.188 & 0.810 & 2.213 & 0.592 & 7736.2 \\

\bottomrule

\end{tabularx}

\end{table*}

Table~\ref{tab:6} summarizes two-dimensional unknown-order performance across 1080 paired scenarios. Compared with NOMP, RC-SAGE--Brent raises order accuracy by 0.369, improves response NMSE by 6.255 dB, and reduces the severe failure rate by 0.248. On the difficult subset of 240 scenarios (\(L\ge3\), SNR \(\ge10\) dB), RC-SAGE--Brent improves order accuracy by 0.796 and NMSE by 7.311 dB, with a paired time ratio of 1.049. Across all 1080 scenarios, the overall paired time ratio rises to 3.939, driven primarily by fixed search overhead in noise-only and simple low-order scenarios.

Stratified results by SNR explain why NOMP degrades in this 100 ms PN benchmark. In nonempty channels, the order accuracies of RC-SAGE--Brent and NOMP are 0.94 versus 0.92 at 0 dB, 0.99 versus 0.50 at 10 dB, and 0.99 versus 0.09 at 20 dB, respectively. Across the same SNR progression, the mean number of spurious paths detected by NOMP escalates from 0.05 to 0.69 and 2.78 per scenario. By contrast, RC-SAGE--Brent holds spurious paths to 0.02, 0.01, and 0.02 per scenario. Thus, the accuracy degradation of NOMP at high SNR stems from order overestimation rather than missed detections. In the 240 noise-only scenarios, RC-SAGE--Brent and NOMP generated one and two false alarms, respectively. The two algorithms also differ in their complex gain updates: RC-SAGE--Brent employs pathwise conditional least squares within SAGE, whereas NOMP refits joint least squares over its active support.

Stress benchmarks with shorter 25 ms probes revealed that the fixed penalty criterion overestimates model order for LFM probes. Allocating additional search iterations does not remove these spurious detections because they originate from structured cancellation residuals along the delay--scale ambiguity ridge. Table~\ref{tab:7} reports production validation results using the residual-adaptive criterion in~\eqref{eq:22b} and~\eqref{eq:22c}.

\begin{table*}[!t]
\caption{Production order validation with 25 ms probes and the LFM ablation}\label{tab:7}
\centering\footnotesize
\setlength{\tabcolsep}{3pt}
\begin{tabularx}{\textwidth}{YYYYYYY}
\toprule

Waveform and order criterion & Scenarios & Order accuracy & Overestimation rate & Underestimation rate & Severe error rate & Median response NMSE/dB \\

\midrule

PN, fixed criterion & 100 & 0.96 & 0.04 & 0 & 0.02 & \ensuremath{-}43.14 \\

LFM, residual-adaptive criterion & 100 & 0.84 & 0.13 & 0.03 & 0.01 & \ensuremath{-}32.51 \\

LFM, fixed-criterion ablation & 100 & 0.51 & 0.49 & 0 & 0.27 & \ensuremath{-}35.00 \\

\bottomrule

\end{tabularx}

\end{table*}

All three configurations satisfied the 15 s deadline, recording maximum output times of 14.80, 14.79, and 14.74 s. The Wilson 95\% score intervals for order accuracy are [90.2\%, 98.4\%] for PN under the fixed criterion, [75.6\%, 89.9\%] for LFM under the adaptive criterion, and [41.3\%, 60.6\%] for LFM under the fixed criterion. On identical LFM candidate sets, the adaptive criterion increases order accuracy by 33 percentage points and decreases the severe error rate from 27\% to 1\%. However, median response NMSE degrades by 2.49 dB, indicating that several weak candidates pruned by the adaptive penalty still contributed to waveform fitting. Across SNR levels, adaptive LFM order accuracy reaches 80\%, 96\%, 80\%, and 80\% at 10, 15, 20, and 25 dB, respectively. Across channel conditions, accuracy reaches 100\% in ordinary, near--far, and parameter-boundary channels, 75\% in dense channels, and 45\% under mixed-stress dynamics. Although this performance meets our 80\% engineering acceptance benchmark, the confidence intervals and stratified results show that reliable order selection across arbitrary propagation regimes remains an open challenge.

We next tested whether improved model-order selection benefits communication decoding. We transmitted independent QPSK multicarrier frames across the 100 frozen LFM validation scenarios. Each scenario transmitted 32 blocks over 33 subcarriers spanning 6--10 kHz, yielding 211200 paired information bits. Both the fixed and adaptive criteria operated on identical received signals, noise realizations, and candidate paths. The receiver reconstructed the delay--scale channel matrix from estimated parameters and applied linear minimum mean-squared error (LMMSE) equalization. The adaptive criterion reduced the bit error rate (BER) from 0.02686 to 0.02578. The mean paired difference across scenarios was \(-0.00108\) (95\% bootstrap confidence interval \([-0.00193,-0.00032]\), two-sided Wilcoxon signed-rank \(p=0.0020\)). Mean error-vector magnitude (EVM) changed by \(+0.011\) dB (\([-0.018,+0.043]\) dB, \(p=0.798\)), and data-operator NMSE changed by \(+0.020\) dB (\([-0.015,+0.064]\) dB, \(p=0.375\)). Neither metric indicated a significant change in continuous response reconstruction. The two criteria selected identical path sets in 59 scenarios. Among the 41 scenarios with differing path sets, the adaptive criterion improved BER in 19 scenarios, tied in 16, and degraded BER in six. Because this simulation assumed perfect synchronization and known noise variance, these findings apply specifically to uncoded hard decisions under linear equalization; verifying end-to-end performance on physical sea trials remains an important direction.

\renewcommand{\arraystretch}{0.94}

\begin{table*}[!t]
\caption{FVB/SVB comparison under a common physical model, matched capacity, and a 15 s deadline}\label{tab:8}
\centering\footnotesize
\setlength{\tabcolsep}{3pt}
\begin{tabularx}{\textwidth}{YYYYYY}
\toprule

Waveform & Method & Order accuracy & Severe error rate & Sparse output NMSE/dB & Full posterior NMSE/dB \\

\midrule

PN & RC-SAGE--Brent & 0.96 & 0.02 & \ensuremath{-}43.14 & \ensuremath{-}43.14 \\

PN & Corrected FVB adaptation & 0.07 & 0.71 & \ensuremath{-}2.86 & \ensuremath{-}2.96 \\

PN & Corrected SVB adaptation & 0.04 & 0.72 & \ensuremath{-}2.56 & \ensuremath{-}2.62 \\

LFM & RC-SAGE--Brent & 0.84 & 0.01 & \ensuremath{-}32.51 & \ensuremath{-}32.51 \\

LFM & Corrected FVB adaptation & 0.10 & 0.61 & \ensuremath{-}12.21 & \ensuremath{-}13.51 \\

LFM & Corrected SVB adaptation & 0.09 & 0.65 & \ensuremath{-}8.87 & \ensuremath{-}9.17 \\

\bottomrule

\end{tabularx}

\end{table*}

\renewcommand{\arraystretch}{1}

Table~\ref{tab:8} compares RC-SAGE--Brent with sample-domain adaptations of FVB and SVB across 100 paired scenarios per waveform. Across all test runs, RC-SAGE--Brent outperforms FVB and SVB in order accuracy by 81.5 and 83.5 percentage points, respectively (exact McNemar test \(p=7.10\times10^{-48}\) and \(4.54\times10^{-49}\)). The median paired difference in sparse-output NMSE (RC-SAGE--Brent minus baseline) is \(-27.83\) dB against FVB (95\% bootstrap interval \([-31.02,-25.49]\) dB) and \(-29.44\) dB against SVB (\([-31.80,-27.81]\) dB). The unpruned VB posterior distributions achieve better fits than their six-path sparse projections, showing that sparse thresholding accounts for part of the performance loss. However, even the full posterior reconstructions remain substantially less accurate than RC-SAGE--Brent.

The 200-scenario VB benchmark ran across four worker processes with one thread per worker. To test whether shared CPU cache or memory bandwidth influenced the runtimes, we reran 24 scenarios sequentially in a single process, stratified by waveform and SNR. The isolated and batch runs yielded identical order estimates in 89.6\% of trials, with a median absolute NMSE difference of only 0.13 dB. In these isolated reruns, RC-SAGE--Brent, FVB, and SVB identified the correct order in 22, one, and two scenarios, respectively, preserving the relative ranking. Multiprocess contention slightly affected the number of iterations completed within the 15 s budget. Table~\ref{tab:8} evaluates sample-domain adaptations of the authors' public algorithms under a fixed execution budget; it does not represent the performance of FVB or SVB in their native multicarrier OFDM environments.

\subsection{Measured-Channel Replay and Communication Performance}\label{sec:7.4}

\begin{figure}[!t]
\centering
\includegraphics[width=\columnwidth,height=.66\textheight,keepaspectratio]{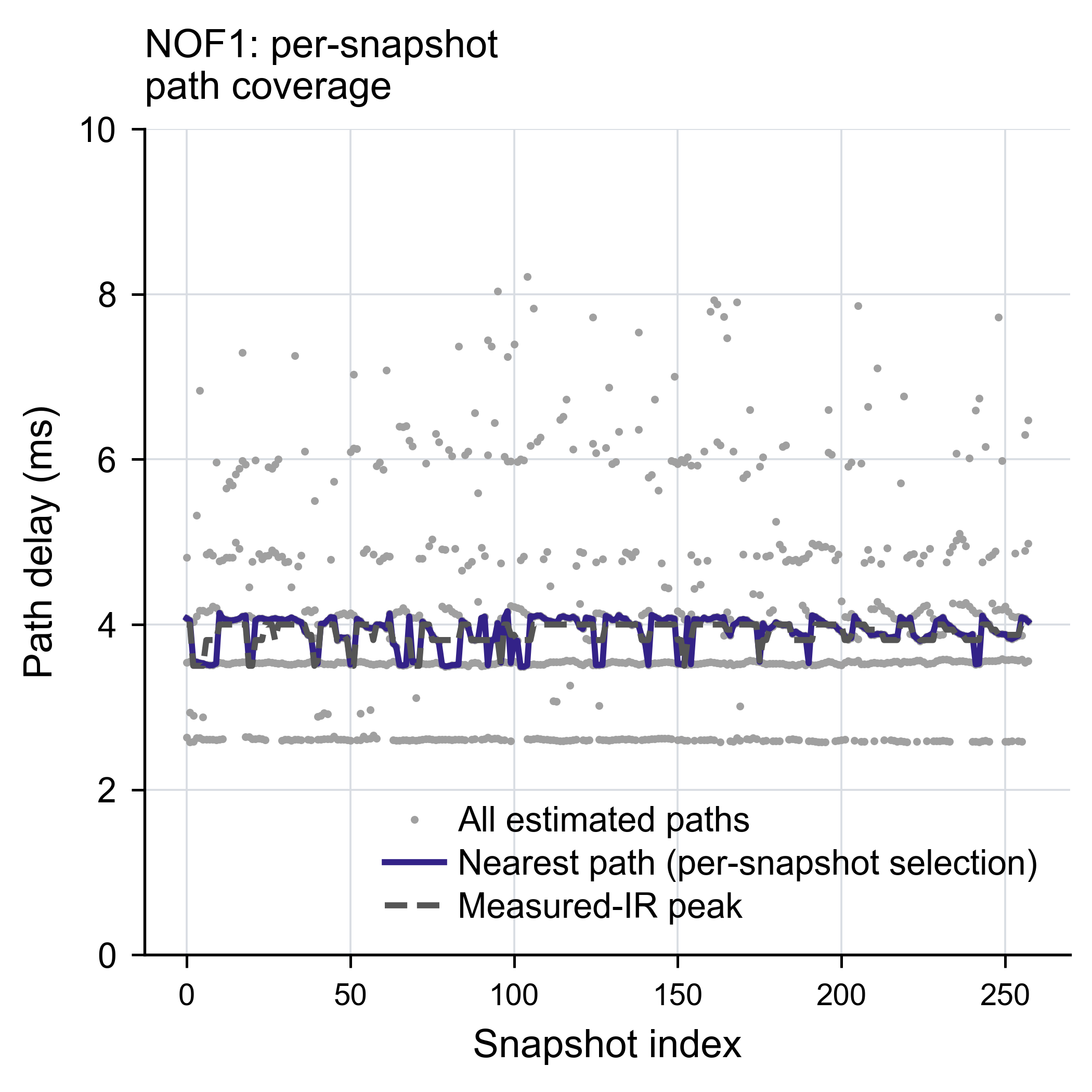}
\caption{Snapshot-wise estimated path delays and measured strongest IR peak for NOF1.}\label{fig:11}

\end{figure}

Fig.~\ref{fig:11} illustrates delay estimates across 258 snapshots of the measured Watermark channel NOF1\_001. Gray markers show all estimated paths. For each snapshot, the red curve traces the estimated arrival closest to the dominant impulse-response peak, marked by the dashed black curve. Across all snapshots, the median absolute delay error is 0.0565 ms, and the 90th percentile is 0.289 ms. We estimate paths independently per snapshot without Kalman tracking or temporal smoothing, confirming that the estimator reliably tracks physical arrival clusters.

\begin{figure*}[!t]
\centering
\includegraphics[width=\textwidth,height=.72\textheight,keepaspectratio]{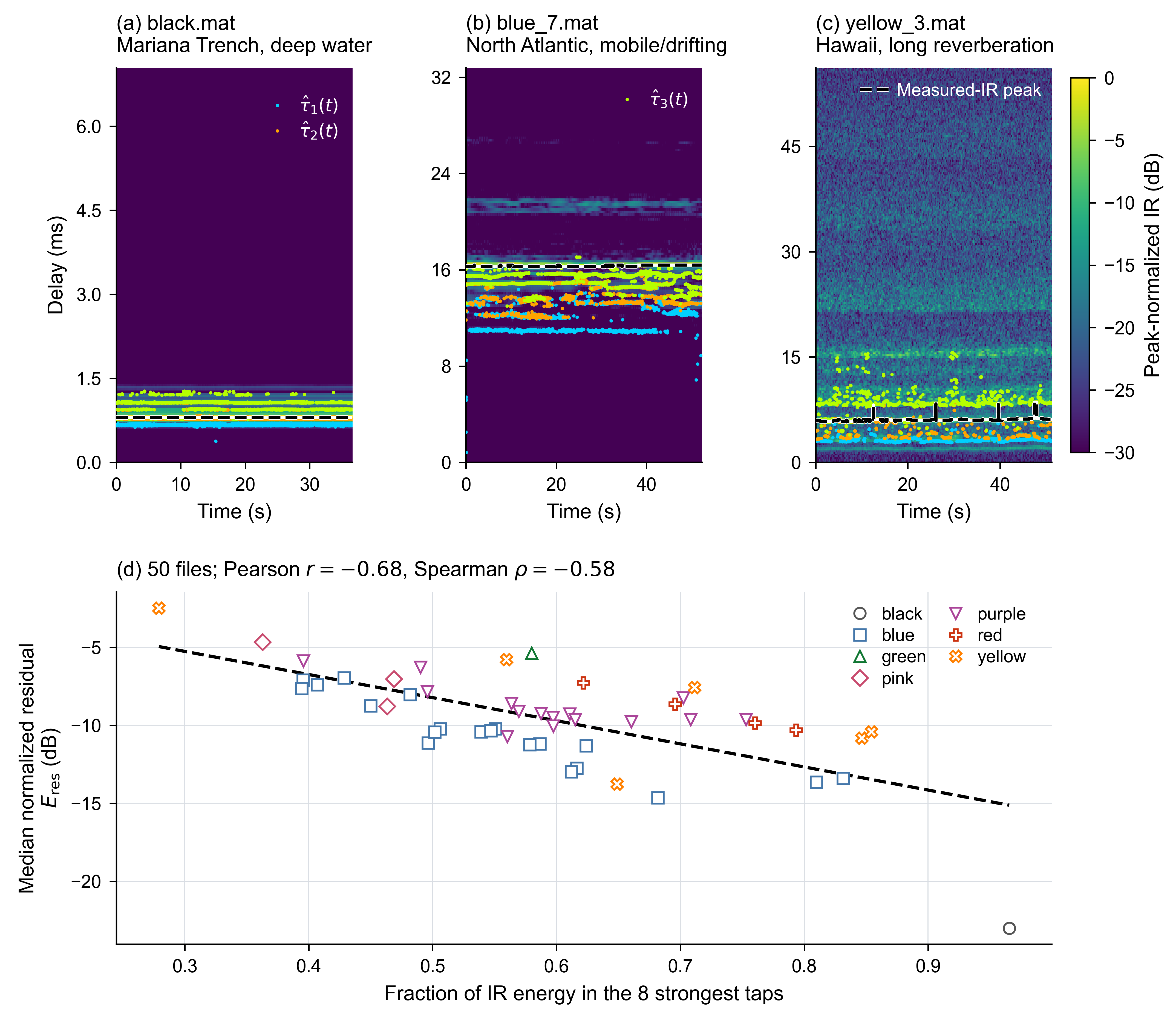}
\caption{Measured UWA IR characteristics: (a) deep-water sparse; (b) mobile drift; (c) long reverberation; (d) energy concentration and fitting residuals for 50 channel records.}\label{fig:12}

\end{figure*}

Figs.~\ref{fig:12}(a)--(c) display deep-water sparse, mobile-drifting, and long-reverberation impulse responses from the UWA Channel Library \cite{Li2026UWALibrary,UWALibrary2026Dataset}. Blue, orange, and green markers denote the three strongest estimated paths in each snapshot, and dark lines trace the measured channel peaks. Fig.~\ref{fig:12}(d) extends this fitting analysis across 50 channel records. Restricting the model to eight paths, the median fitting residual \(E_{\mathrm{res}}\) in~\eqref{eq:8a} spans \(-23.00\) dB to \(-2.49\) dB across records, with a cross-record median of \(-9.62\) dB. The energy fraction contained within the eight strongest physical taps correlates with \(E_{\mathrm{res}}\) at \(-0.68\) (Pearson) and \(-0.58\) (Spearman). Because \(E_{\mathrm{res}}\) reflects both measurement noise and diffuse scattering, this correlation shows that energy concentration predicts how well a sparse multipath model captures a physical channel.

Communication replay evaluates the Watermark \cite{vanWalree2017Watermark} channels NOF1\_001 and KAU1\_001 at SNRs of 10, 20, and 30 dB, using 30 paired frames per SNR level. Under the snapshot grouping described in Section~\ref{sec:6.1}, an ideal equalizer with access to the full channel response (Perfect-dense) achieves three-SNR mean BERs of 0.0491 and 0.0744 on the two channels. The true channels contain 94.8\% and 81.8\% of their total energy within the cyclic prefix. When the equalizer retains only the 10 strongest true channel taps, the mean BER increases to 0.1618 and 0.1855, respectively. Both diffuse multipath tails and energy leaking past the cyclic prefix degrade equalization performance.

The replay synthesizes received waveforms by convolving transmitted signals with complex baseband impulse responses from each snapshot, rather than synthesizing analytic bandpass responses via~\eqref{eq:1}. Expressing the analytic bandpass signal as \(s(t)=u(t)e^{j2\pi f_ct}\), downconverting the time-scaled path \(\ell\) with an oscillator at \(f_c\) yields the baseband envelope

\begin{equation}\tag{30}\label{eq:30}
\fitmath{\begin{gathered}
\widetilde\alpha_\ell\,
u\!\left((1+\epsilon_\ell)(t-\tau_\ell)\right)
e^{j2\pi f_c\epsilon_\ell t},\\
\widetilde\alpha_\ell=
\alpha_\ell e^{-j2\pi f_c(1+\epsilon_\ell)\tau_\ell}.
\end{gathered}}
\end{equation}

Although the constant phase offset can be absorbed into the complex path gain, the time-varying Doppler carrier phase must be tracked explicitly. Because the Watermark impulse responses remain static within each transmission frame, our replay evaluates delay multipath estimation without Doppler scale estimation.

\begin{figure*}[!t]
\centering
\includegraphics[width=\textwidth,height=.72\textheight,keepaspectratio]{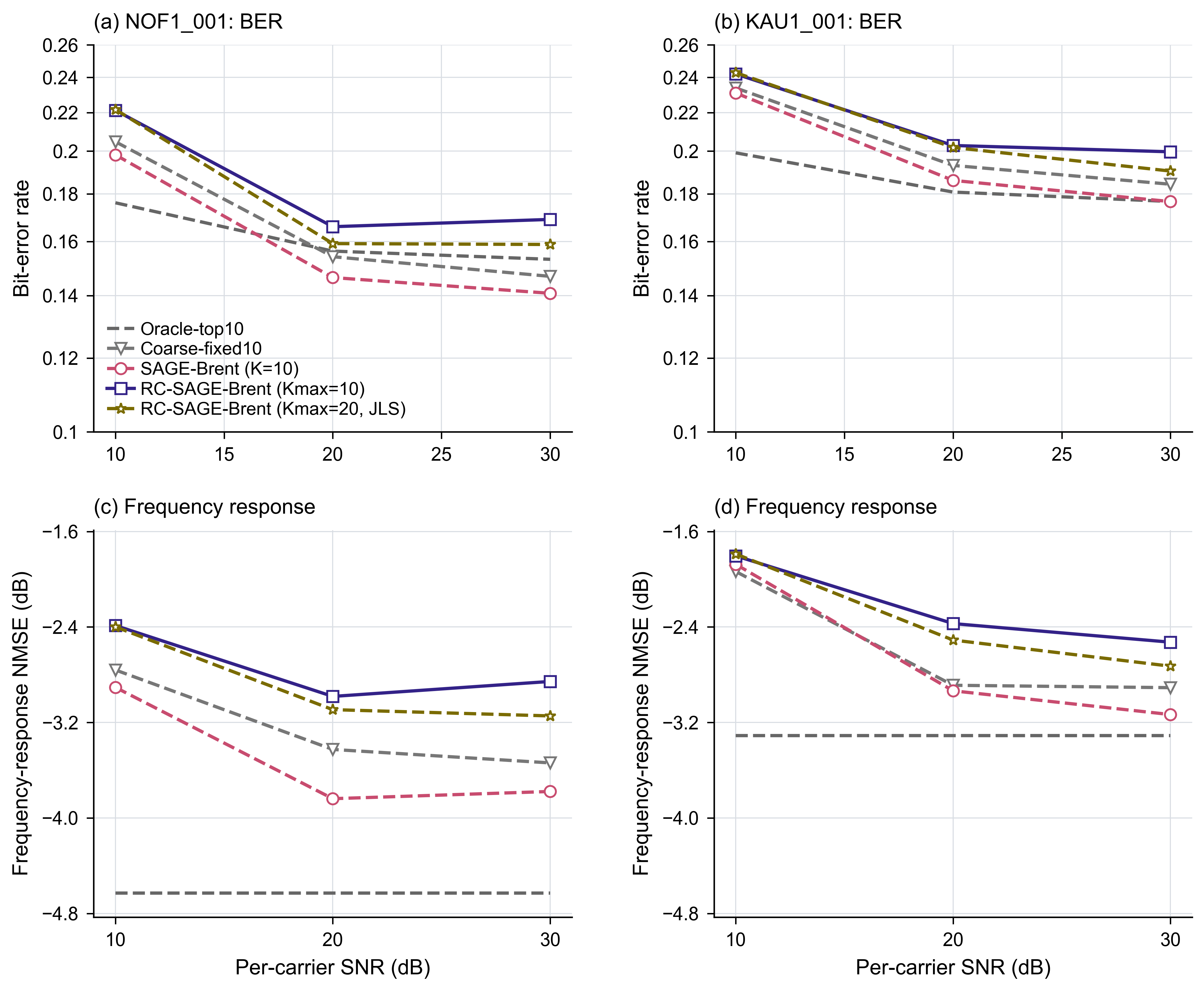}
\caption{Watermark IR replay: (a)(b) NOF1 and KAU1 BER; (c)(d) corresponding frequency-response NMSE.}\label{fig:13}

\end{figure*}

\begin{table*}[!t]
\caption{Paired BER comparisons grouped by snapshot and summarized within channels}\label{tab:9}
\centering\footnotesize
\setlength{\tabcolsep}{3pt}
\begin{tabularx}{\textwidth}{YYYYY}
\toprule

Comparison & BER reduction & EVM gain/dB & Frequency-response NMSE gain/dB & Path-count change \\

\midrule

Continuous-fixed10 versus Coarse-fixed10 & 0.0063 & 0.29 & 0.17 & 0 \\

RC-CFAR10 versus Continuous-fixed10 & \ensuremath{-}0.0203 & \ensuremath{-}1.09 & \ensuremath{-}0.59 & \ensuremath{-}2.63 \\

CFAR20-all versus CFAR10 & 0.0045 & 0.18 & 0.12 & +2.82 \\

CFAR20-strong10 versus CFAR20-all & \ensuremath{-}0.0027 & \ensuremath{-}0.08 & \ensuremath{-}0.03 & \ensuremath{-}1.88 \\

\bottomrule

\end{tabularx}

\end{table*}

Each point in Fig.~\ref{fig:13} uses the 30 snapshots of the corresponding channel, and the three SNRs share the same snapshots. Perfect-dense uses the true channel with all taps. Oracle-top10 keeps the ten strongest true taps, which separates receiver limitations from sparse-representation error.

Table~\ref{tab:9} reports paired communication differences averaged across the three SNRs within each snapshot and summarized equally across both channels. For a fixed model order of ten paths, continuous SAGE--Brent refinement reduces the mean BER by 0.0063 compared with coarse-grid estimation. When estimating order dynamically, RC-CFAR10 selects 2.63 fewer paths on average, which increases the BER by 0.0203 relative to the 10-path continuous baseline. Expanding the path budget to 20 paths recovers part of this loss, whereas truncating the estimates to the ten strongest paths degrades decoding performance again. Because gain reestimation schemes are held constant across controls, these comparisons isolate the effect of path-set selection. Retaining too few multipath components degrades channel equalization, even when the retained paths are localized accurately. These conclusions apply specifically to the recording conditions of NOF1 and KAU1. Furthermore, the replay highlights that sparse path detection and multicarrier equalization serve distinct operational objectives: diffuse energy tails, while negligible in path detection, contribute constructively to frequency-domain channel inversion.

\subsection{Computational Complexity and Runtime}\label{sec:7.5}

\begin{figure*}[!t]
\centering
\includegraphics[width=\textwidth,height=.72\textheight,keepaspectratio]{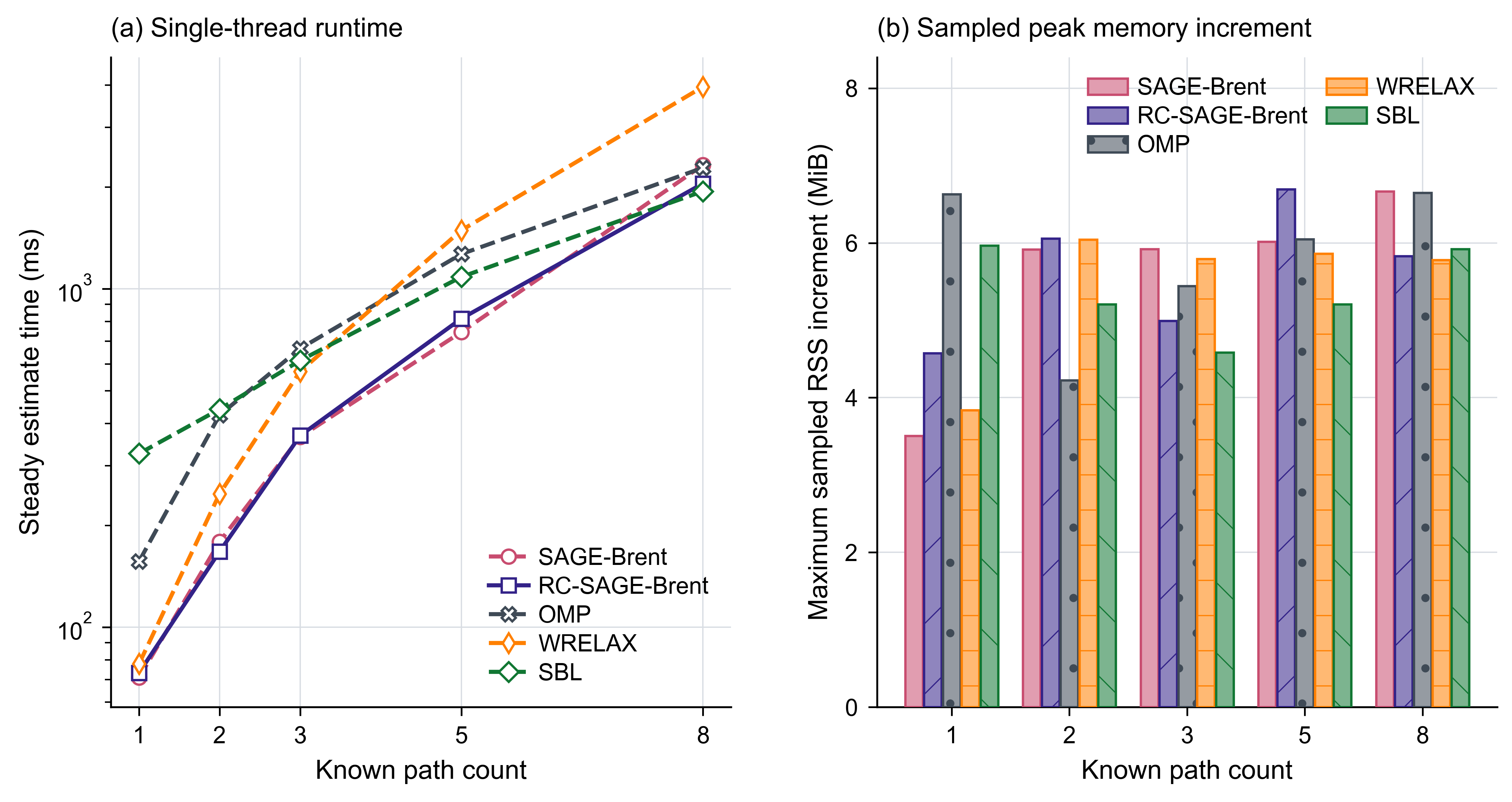}
\caption{Single-thread repeated-estimation cost: (a) median time; (b) maximum sampled memory increment.}\label{fig:14}

\end{figure*}

Fig.~\ref{fig:14} benchmarks computational complexity under known model order using a 25 ms LFM probe at 10 dB SNR. We record 10 repeated estimation calls per method and path count, following one warm-up call. Fig.~\ref{fig:14}(a) reports median execution time on a logarithmic scale, excluding waveform precomputation and process initialization, with all linear algebra libraries pinned to one thread. Fig.~\ref{fig:14}(b) presents the maximum resident set size (RSS) increment recorded at 2 ms sampling intervals relative to the memory baseline prior to execution. This increment excludes library runtimes and static dictionary storage. Section~\ref{sec:5.3} and Appendix~\ref{app:C} provide asymptotic memory derivations.

Fig.~\ref{fig:14}(a) shows execution time scaling with path count \(L\) across five estimators; Supplementary Table S4 lists all numerical measurements. At \(L=1\), standalone SAGE--Brent and RC-SAGE--Brent require 71.0 ms and 73.3 ms, respectively. At \(L=8\), their runtimes increase to 2322.5 ms and 2048.3 ms, whereas OMP, WRELAX, and SBL require 2279.7 ms, 3957.3 ms, and 1944.0 ms, respectively.

Across all 50 paired repeated calls, the median runtime ratios of RC-SAGE--Brent relative to SAGE--Brent, OMP, WRELAX, and SBL are 1.037, 0.557, 0.668, and 0.642, respectively. For \(L=8\), the 10th-to-90th percentile ranges encompass 1.0 in comparisons against OMP and SBL. The peak memory footprint of RC-SAGE--Brent spans 84.6--86.2 MiB, with a maximum stage increment of 6.7 MiB, virtually identical to standalone SAGE--Brent (84.5--86.2 MiB, 6.7 MiB increment). For every evaluated method, the maximum measured memory increment remains below 8 MiB.

We profiled optional near--far replacement separately under identical single-thread conditions. In a two-path near--far scenario (25 dB NFR, \(1.3W_\tau\) spacing), median runtimes without and with replacement are 1197.4 ms and 1066.3 ms, yielding a paired ratio of 0.870 [0.818, 0.995]. In random multipath channels with \(L=5\) and \(L=8\), the corresponding ratios remain at 1.004 [0.957, 1.026] and 0.993 [0.971, 1.108], respectively (Supplementary Table S5). In benign channels, runtime and memory remain virtually unchanged. In near--far channels, runtime decreases because replacing a false component accelerates outer-loop convergence. The replacement check itself introduces only an \(\mathcal O(L)\) dynamic-range test when inactive. When activated, rebuilding the residual, refining one delay coordinate, and recomputing the residual sum of squares requires \(\mathcal O((L+K_\tau)N)\) time and \(\mathcal O(N)\) temporary memory. Because the algorithm accepts at most one replacement per trial, this routine does not alter the asymptotic complexity of the estimator.

We also measured the PR simplification across single-thread runs in isolated processes. The median paired PR-to-VPR runtime ratios for one-dimensional \(L=5\), one-dimensional \(L=8\), and two-dimensional \(L=5\) are 1.036, 0.964, and 0.984, respectively. Peak memory differs by at most 1.0 MiB between the two configurations. Omitting candidate validation alters runtime by only 1\%--4\%, while retaining validation preserves weak-path detectability under extreme near--far conditions (Table~\ref{tab:3}).

In summary, RC-SAGE--Brent and standalone SAGE--Brent exhibit comparable runtimes and memory demands. Aggregate paired runtime ratios favor RC-SAGE--Brent over OMP, WRELAX, and SBL in this known-order benchmark. Unknown-order estimation incurs additional overhead for coarse discovery and redetection, which we report separately. Residual-adaptive order selection reuses least-squares quantities from the fixed criterion, requiring no extra candidate generation or Brent searches. In paired evaluations across 100 scenarios, fixed and adaptive selectors required 0.26155 s and 0.26171 s on average. In the production validation suite, adaptive LFM order selection averaged 0.300 s, with every trial completing within the 15 s budget.

\section{Discussion}\label{sec:8}

\subsection{Roles of Estimation Steps and Related Methods}\label{sec:8.1}

The RC-SAGE--Brent architecture addresses continuous parameter errors directly. Off-grid parameter offsets leave structured cancellation residuals after strong-path subtraction. Continuous alternating SAGE--Brent searches suppress these residuals, and cyclic conditional least-squares refitting updates the residual representation for the full active support. Candidate validation and resolution pruning then leverage the cleaned residual to eliminate false peaks. Finally, full-domain redetection identifies genuine weak arrivals previously obscured by strong-path energy. Any structural modification to the path set initiates another refinement cycle. This closed feedback loop allows the estimator to correct early misidentifications that would otherwise corrupt sequential forward estimators. Three lines of evidence validate this mechanism: the analytical cancellation identities in~\eqref{eq:24}--\eqref{eq:28}, the factorial ablations demonstrating the complementary roles of pruning and redetection, and the two-dimensional benchmarks showing substantial NMSE and recall gains. While candidate validation and resolution pruning perform redundant roles in standard random channels, validation provides indispensable protection in near--far boundary regimes by preventing false peaks from exhausting the path budget.

Our framework builds upon the insight of robust OFDM-SAGE \cite{PintoJuntti2026RobustOFDMSAGE}, which demonstrated the value of unifying parameter updates and order determination. We extend this principle to wideband delay--scale estimation by incorporating ambiguity-domain search constraints and bidirectional path revision. These domain constraints govern how cancellation residuals form and ensure that newly exposed residuals are categorized properly as new arrivals rather than model artifacts.

Other contemporary approaches leverage alternative structures. Deep-learning estimators \cite{FengEtAl2024UncertainSparsity} rely on offline training corpora. The time-domain off-grid SBL method of Yang and Choo \cite{YangChoo2025TimeDomainOGSBL} pools spatial information across repeated stationary transmissions. Underwater OTFS formulations \cite{YangEtAl2026UWAOTFS} exploit localized Doppler delay-chirp geometries. In contrast, our approach operates directly on a single wideband probe transmission, refining continuous delays and Doppler scales simultaneously without requiring repeated snapshots or prior training sets.

In the one-dimensional difficult test suite, RC-SAGE--Brent achieved superior order accuracy and lower reconstruction error than OMP, generated fewer severe overestimation errors than WRELAX, and retained more weak arrivals than SBL under identical dictionary and CFAR stopping constraints. In two-dimensional known-order benchmarks, RC-SAGE--Brent and NOMP exhibited comparable recall in benign channels, with RC-SAGE--Brent achieving substantial NMSE improvements and superior near--far recovery. In 25 ms LFM production tests, the residual-adaptive penalty increased order accuracy from 51\% to 84\% on identical candidates, while reducing severe order errors from 27\% to 1\%. However, median response NMSE degraded by 2.49 dB, and order accuracy under mixed-stress conditions remained at 45\%. The fixed coarse-grid search overhead increased total execution time in simple noise-only channels, while achieving a paired runtime ratio of 1.049 relative to NOMP on the difficult subset.

\subsection{Applicability Limits and Engineering Choices}\label{sec:8.2}

We identify five operational boundaries of the proposed method:

\begin{enumerate}

\item \textbf{High model orders under low SNR}: At \(L=8\) and 0 dB SNR, all evaluated algorithms encounter fundamental detectability limits, with RC-SAGE--Brent achieving an order accuracy of 0.25.

\item \textbf{Subresolution path separations}: When path arrivals lie within the main-lobe resolution limit, correlation peaks coalesce and parameter identifiability degrades. Resolution pruning then merges closely spaced arrivals into a single effective component.

\item \textbf{Capacity saturation in near--far channels}: In known-order trials where strong-path residual peaks fill the candidate budget early, genuine weak paths cannot enter the active set. The optional replacement routine resolves this specific scenario by testing path substitutions, though standard unknown-order estimation omits this step.

\item \textbf{Delay--scale ambiguity coupling}: The shearing ridge of the LFM ambiguity function degrades the conditioning of the local curvature matrix \(\mathbf Q\), making alternating optimization more sensitive to initialization than under PN probes. The residual-adaptive penalty improves LFM order accuracy to 84\% and reduces communication BER by 0.00108, but it degrades median response NMSE by 2.49 dB. This tradeoff between order robustness and waveform fitting highlights the difficulty of joint delay--scale estimation under highly coupled probes.

\item \textbf{Diffuse multipath energy tails}: In physical channels characterized by extended reverberation tails, individual scattered arrivals often fall below CFAR detection thresholds. Although these components cannot be resolved as discrete paths, they contribute constructively to equalization, meaning that sparse order selection may underestimate the tap count needed for optimal communication decoding.

\end{enumerate}

These operational limits guide practical deployment. For physical localization, channel tracking, and sparse arrival estimation, full RC-SAGE--Brent with VPR filtering provides robust continuous resolution and automated order selection. When equalizing channels with dense reverberant tails, practitioners should combine discrete sparse estimation with a diffuse tail model or optimize the effective tap count directly against equalizer performance metrics.

\section{Conclusion}\label{sec:9}

We developed RC-SAGE--Brent, a residual-closed-loop estimator that couples continuous delay--scale refinement with unknown-order path-set correction. Ambiguity-constrained alternating Brent searches and cyclic conditional updates suppress strong-path cancellation residuals. Candidate validation, resolution-constrained pruning, and full-domain residual redetection then revise the active path set, allowing the estimator to correct early false detections and recover masked weak arrivals. Our mathematical analysis links off-grid parameter offsets directly to cancellation-residual curvature and weak-path detectability.

Ablation experiments confirm that pruning stabilizes model order, redetection recovers obscured arrivals, and candidate validation prevents premature capacity saturation under severe near--far dynamics. In synthetic benchmarks against OMP, WRELAX, SBL, and two-dimensional NOMP, RC-SAGE--Brent demonstrates superior reconstruction accuracy and robust model-order estimation across both PN and LFM probes. Under a strict 15 s real-time budget, the residual-adaptive criterion achieved 84\% order accuracy for 25 ms LFM probes, mitigating overestimation caused by delay--scale coupling. Measured channel replays demonstrate reliable physical path tracking while illustrating the distinction between sparse arrival detection and dense equalization requirements. Future work will investigate joint parameterizations of sparse arrivals and diffuse reverberation tails to enhance communication performance in reverberant underwater channels.

\appendices

\section{Proof of the Cancellation-Residual Energy Identity and Local Curvature}\label{app:A}

Let \(\mathbf P_\delta=\mathbf a_\delta\mathbf a_\delta^{\mathsf H}/\|\mathbf a_\delta\|_2^2\). The least-squares fitted vector is \(\widehat{\mathbf x}=\mathbf P_\delta\mathbf x\), hence

\begin{equation}\tag{A.1}\label{eq:A.1}
\fitmath{\frac{\|\mathbf x-\widehat{\mathbf x}\|_2^2}{\|\mathbf x\|_2^2}
=\frac{\mathbf a_0^{\mathsf H}(\mathbf I-\mathbf P_\delta)\mathbf a_0}
{\|\mathbf a_0\|_2^2}
=1-\frac{|\mathbf a_\delta^{\mathsf H}\mathbf a_0|^2}
{\|\mathbf a_\delta\|_2^2\|\mathbf a_0\|_2^2},}
\end{equation}

which is~\eqref{eq:24}. Expand the single-path response at \(\boldsymbol\theta_0\) as

\(\mathbf a_\delta=\mathbf a_0+\mathbf D\boldsymbol\delta+ \mathcal O(\|\boldsymbol\delta\|^2)\). Since the first-order change along \(\mathbf a_0\) is absorbed by the unknown complex gain, projection onto the orthogonal complement of the true single-path response leaves the lowest-order nonzero residual term

\(\mathbf P_\perp\mathbf D\boldsymbol\delta\). Dividing by \(E_a\) and taking the real quadratic form gives~\eqref{eq:25}.

\section{Curvature--Fisher Information Relationship}\label{app:B}

Take the parameter vector

\(\boldsymbol\xi=[\operatorname{Re}\alpha, \operatorname{Im}\alpha,\tau,\epsilon]^{\mathsf T}\).

The Fisher information of the complex Gaussian mean model is

\begin{equation}\tag{B.1}\label{eq:B.1}
\fitmath{[\mathbf J]_{ij}=\frac{2}{\sigma^2}
\operatorname{Re}\left\{
\frac{\partial\boldsymbol\mu^{\mathsf H}}{\partial\xi_i}
\frac{\partial\boldsymbol\mu}{\partial\xi_j}\right\}.}
\end{equation}

Taking the Schur complement of the gain block is equivalent to projecting \(\alpha\mathbf D\) onto the orthogonal complement of

\(\operatorname{span}(\mathbf a_0)\), yielding

\begin{equation}\tag{B.2}\label{eq:B.2}
\fitmath{\mathbf J_{\tau,\epsilon\mid\alpha}
=\frac{2|\alpha|^2}{\sigma^2}
\operatorname{Re}\{\mathbf D^{\mathsf H}
\mathbf P_\perp\mathbf D\},}
\end{equation}

The definition of \(\mathbf Q\) in~\eqref{eq:25} then gives~\eqref{eq:26}. If \(\mathbf Q\) is nonsingular, the local CRB is the inverse of this information submatrix; near singularity increases the variance lower bound in the corresponding parameter direction.

\section{Update Properties and Objective-Evaluation Complexity}\label{app:C}

With scale fixed, step (15) accepts only delay updates that do not decrease \(J_\ell\); step (16) does the same for scale with the new delay fixed. For given parameters,~\eqref{eq:13} minimizes squared residuals. A full delay--scale--gain cycle therefore does not reduce the current single-path objective. It is bounded by \(\|\mathbf z_\ell\|_2^2\), so the accepted objective-value sequence converges.

A dense 2-D grid requires \(N_\tau N_\epsilon\) objective evaluations for one search over the entire range. If Brent uses \(q_\tau,q_\epsilon\) evaluations on the axes, with \(K_{\mathrm{in}}\) inner iterations, \(K_{\mathrm{out}}\) outer SAGE cycles, and \(L\) paths, the scalar objective-evaluation count of continuous refinement scales as

\begin{equation}\tag{C.1}\label{eq:C.1}
\fitmath{\mathcal O\!\left(
K_{\mathrm{out}}L K_{\mathrm{in}}(q_\tau+q_\epsilon)
\right).}
\end{equation}

Unknown-order RC-SAGE--Brent also requires coarse-grid search, pathwise conditional LS cycles, pruning, and redetection; 2-D estimation additionally searches scale in one dimension. Both 1-D and 2-D estimators use pathwise conditional LS and cache computed single-path responses to avoid repeated related computations. Actual runtime is jointly determined by evaluations in~\eqref{eq:C.1}, response generation, and linear algebra.

For optional known-order near--far replacement in Section~\ref{sec:7.1}, excluded from the standard scheme, a below-threshold estimated-gain dynamic range requires only an \(\mathcal O(L)\) magnitude-range check. A replacement test forms the fitting residual with the weakest component removed, performs one delay Brent refinement, generates a candidate response, and compares SSE before and after replacement. For observation length \(N\) and \(K_\tau\) Brent evaluations, the additional time is

\begin{equation}\tag{C.2}\label{eq:C.2}
\fitmath{\begin{gathered}
\mathcal O\!\left((L+K_\tau)N\right),
\\
\text{additional working memory}=\mathcal O(N).
\end{gathered}}
\end{equation}

At most one replacement is accepted per scenario, so this extension does not change the main search's asymptotic order in~\eqref{eq:C.1}. Supplementary Table S5 gives measured runtime.

\section*{Reproducibility and Supplementary Material}

Method parameters, panel-wise experimental settings, baseline evaluation budgets, and repeat-estimation details are provided in the accompanying supplementary material.

\bibliographystyle{IEEEtran}

\bibliography{references_verified}

\clearpage
\onecolumn
\section*{Supplementary Material}

Tables S1--S5 document the method configurations, protocols, and timing details of the original experiments. The main text describes the residual-adaptive order criterion and reports the additional 25 ms probe experiments and FVB/SVB comparisons.

\subsection*{Table S1. Method configurations in the main text}
\noindent\begin{tabularx}{\linewidth}{lYYYY}\toprule
Method & Continuous SAGE--Brent & Validation/pruning & Global residual redetection & Purpose\\\midrule
SAGE--Brent & Yes & No & No & Ablation configuration of the proposed refinement module\\
RC-SAGE--Brent & Yes & Yes & Yes & Proposed unknown-order method\\\bottomrule
\end{tabularx}

\subsection*{Table S2. Figure 10 panels and protocols}
\noindent\begin{tabularx}{\linewidth}{p{1.6cm}YYY}\toprule
Panel & Scenario and budget & Samples and pairing unit & Supported conclusions\\\midrule
(a) Near--far diagnostic & PN, two known paths, overall SNR 25 dB; NFR 15/25/30/35 dB; scale separation 0 or \(0.4W_\epsilon\) & 20 base scenarios per scale separation; 40 base scenarios repeated at four NFRs yield 160 paired conditions; 40 records per method at each NFR & Joint location/amplitude recovery under specified diagnostic conditions; does not represent overall recall or unknown-order performance\\
(b) Random SNR sweep & PN, known \(L=3,5\), SNR 0/10/20/30 dB & 20 independent scenarios per cell, 160 scenarios and 320 method records; methods paired within scenario & Response NMSE trends within the tested path counts and SNRs; does not independently identify the cause of an error floor\\
(c)--(h) Unknown order & \(L=0,1,2,3,5,8\), SNR 0/10/20 dB, \(L_{\max}=10\) & 80/80/80/60/40/20 scenarios in the six order cells per SNR, totaling 1080 scenarios and 2160 method records; both rows share data & Order selection and spurious-path control; limits of 0 dB and high-order cells are retained\\\bottomrule
\end{tabularx}

\subsection*{Table S3. Baseline implementations and stopping budgets}
\noindent\begin{tabularx}{\linewidth}{p{1.8cm}YYY}\toprule
Method & Candidates and continuous updates & Gain updates & Main stopping conditions/budgets\\\midrule
RC-SAGE--Brent & Coarse discovery of a list followed by Brent; sorted validation, pruning, redetection & SAGE pathwise conditional LS cycles & Order increases limited by threshold and \(L_{\max}\); outer loop stops at coordinate/support stability or 12 cycles\\
OMP & Strongest residual coarse peak; single-path Gauss--Newton & Explicit joint LS after every admission & Next peak below CFAR; at most 60 evaluations per path and one final refinement pass\\
WRELAX & Global 1-D ML initialization; cyclic Brent relaxation & Final explicit joint LS & Next peak below CFAR; at most 8 relaxation passes or movement at most \(10^{-3}/f_s\)\\
SBL & 64 candidates spaced at \(T_s/4\); 4 Brent passes on retained support & Explicit joint LS after each admission and at the end & At most 200 EM iterations; relative hyperparameter change below \(10^{-2}\) after 10 iterations, or next support below CFAR\\\bottomrule
\end{tabularx}

\subsection*{Table S4. Single-thread median repeat-estimation time (ms)}
\noindent\begin{tabularx}{\linewidth}{Yrrrrr}\toprule
Method & \(L=1\)&\(L=2\)&\(L=3\)&\(L=5\)&\(L=8\)\\\midrule
SAGE--Brent &71.0&179.0&365.2&745.1&2322.5\\
RC-SAGE--Brent &73.3&167.4&368.4&816.6&2048.3\\
OMP &156.5&423.8&666.3&1267.2&2279.7\\
WRELAX &78.0&248.0&570.5&1489.5&3957.3\\
SBL &326.0&441.0&614.3&1084.7&1944.0\\\bottomrule
\end{tabularx}

\subsection*{Table S5. Single-thread repeated estimation with optional near--far recovery}
The near--far scenario uses NFR 25 dB, spacing \(1.3W_\tau\), and \(L=2\); both random scenarios use SNR 10 dB.

\noindent\begin{tabularx}{\linewidth}{YYYYY}\toprule
Scenario & Median time: main/recovery (ms) & Time ratio [P10,P90] & Evaluations: main/recovery & Absolute peak RSS: main/recovery (MiB)\\\midrule
Near--far \(L=2\)&1197.4 / 1066.3&0.870 [0.818,0.995]&13727.6 / 13438.7&85.9 / 86.0\\
Random \(L=5\)&843.3 / 846.8&1.004 [0.957,1.026]&9244.8 / 9244.8&86.1 / 86.2\\
Random \(L=8\)&2484.5 / 2430.0&0.993 [0.971,1.108]&15698.8 / 15706.8&85.9 / 86.0\\\bottomrule
\end{tabularx}

\subsection*{Reference verification and comparison scope}
The Yang--Choo reference is \emph{Time-Domain Off-Grid Sparse Bayesian Learning for High-Resolution Time-Delay Estimation}, IEEE Access 13 (2025), 170372--170379, DOI 10.1109/ACCESS.2025.3614348. No local Zotero record was found. The authors' institutional publication information and the publicly available full paper were checked. Sections II--III of that paper use a matched-filtered time-domain linear model, first-order delay offsets, and multisnapshot Bayesian inference; Section IV constructs an MMV model using SAVEX15 repeated transmissions under approximate stationarity. We retain it as related work, do not describe the present SBL/SBL-MP curves as a reproduction of that algorithm, do not interpret ``time domain'' as overcoming all physical bandwidth limits, and do not directly compare runtime across hardware.

\subsection*{Figure scope and retained data}
\begin{itemize}
\item Figure 1 omits the original performance panel (d), retaining the fixed-grid error floor and mismatch leakage on both axes; weak-path performance is discussed in Figure 7 and the ablations.
\item The frozen, identically configured single-path 2-D sweep in Figures 4(b)(c) has no SBL/OGSBL curves. Two-dimensional SBL data from other experiments cannot be inserted directly into this SNR sweep; adding curves later requires separate authorization and a paired protocol.
\item Coarse has been restored in Figure 5's objective-evaluation panels following subsequent requirements; Figure 14's repeat-estimation time and memory still omit Coarse. Original records are retained. The repeat-estimation source contains 300 records for six methods, of which Figure 14 uses 250 records for five methods; the pairing unit remains 50 calls per method.
\item The Figure 6(a) inset is restored to the original upper-right position, copying coordinates from the original curves and showing only the vicinity of 10 dB, with no new simulation points. Figure 10 panels are individually labeled (a)--(h).
\end{itemize}

\subsection*{Point-estimate presentation}
The main text and figures display only point estimates while retaining within-scenario pairing and BER snapshot-grouped summaries. P10 and P90 in Table S5 are empirical quantiles of repeated-call time ratios. Experimental source records are unchanged.

\end{document}